\documentclass[10pt]{amsart}
\usepackage{amsthm}
\usepackage{amsmath}
\usepackage[a4paper, margin=0.75in]{geometry}
\usepackage{hyperref}
\usepackage{graphicx}
\usepackage{comment}
\usepackage{cite}
\DeclareMathOperator{\arcsinh}{arcsinh}
\usepackage{xcolor}
\usepackage{bm}
\DeclareMathOperator\erf{erf}
\DeclareMathOperator\Tr{Tr}
\newcommand{\Jss}{J_{\text{ss}}}

\newcommand{\bp}{\mathbf{p}}

\newcommand{\argmin}{\text{argmin}}

\newtheorem{remark}{Remark}[section]  

\colorlet{red}{black}

\title{Coarse-graining nonequilibrium diffusions with Markov chains}

\author[R. Nartallo-Kaluarachchi]{Ramón Nartallo-Kaluarachchi$^{*,\dagger}$}
\address{$^*$Mathematical Institute, University of Oxford, Woodstock Road, Oxford, OX2 6GG, United Kingdom}
\address{$^\dagger$Centre for Eudaimonia and Human Flourishing, University of Oxford, 7 Stoke Pl, Oxford, OX3 9BX, United Kingdom}
\email{nartallokalu@maths.ox.ac.uk}
\author[R. Lambiotte]{Renaud Lambiotte$^{*}$}
\author[A. Goriely]{Alain Goriely$^{*}$}

\begin{document}
\begin{abstract}
We investigate nonequilibrium steady-state dynamics in both continuous- and discrete-state stochastic processes. Our analysis focuses on planar diffusion dynamics and their coarse-grained approximations by discrete-state Markov chains. Using finite-volume approximations, we derive an approximate master equation directly from the underlying diffusion and show that this discretisation preserves key features of the nonequilibrium steady-state. In particular, we show that the entropy production rate of the approximation converges as the number of discrete states goes to the limit. These results are illustrated with analytically solvable diffusions and numerical experiments on nonlinear processes, demonstrating how this approach can be used to explore the dependence of the entropy production rate on model parameters. Finally, we address the problem of inferring discrete-state Markov models from continuous stochastic trajectories. We show that discrete-state models significantly underestimate the true entropy production rate. However, we also show that they can provide tests to determine if a stationary planar diffusion is out of equilibrium. This property is illustrated with both simulated data and empirical trajectories from schooling fish.
\end{abstract}
\maketitle

\section{Introduction}
\label{sec: intro}
Many physical processes can be described by the interplay of deterministic forcing and random fluctuations. Such processes are typically described as \textit{diffusive} and have been used to model a plethora of systems including Brownian motion \cite{brown1828brownian}, cell motility \cite{Li2011dicty}, climate dynamics \cite{Franzke_O’Kane_2017}, neural systems \cite{harrison2005stochasticneuronal}, and financial markets, amongst others. Mathematically, they can be formulated as Itô \textit{stochastic differential equations} (SDE), taking the form
\begin{align}
    dX(t) = f(X(t))\;dt + \Sigma(X(t))\;dW(t),\label{eq: SDE} 
\end{align}
where $X(t)$ represents the state of the process at time $t$, $f(x)$ represents the deterministic forcing, and $\Sigma(x)$ determines the random fluctuations as $W(t)$ is the standard \textit{Wiener process} \cite{pavliotis2014stochproc}.\\
\begin{figure}
    \centering
    \includegraphics[width=\linewidth]{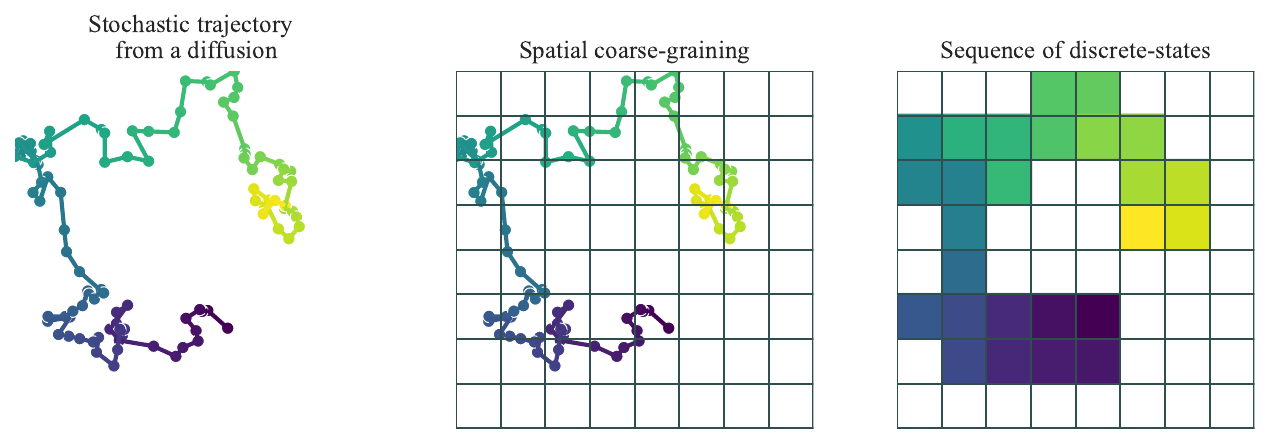}
    \caption{\textbf{Coarse-graining a stochastic trajectory}. A stochastic trajectory from a diffusion can be modelled as a sequence of discrete states by performing a discretisation of state space.}
    \label{fig: traj coarsegrain}
\end{figure}
\\
It has been argued by Schr\"odinger that biological systems  stave off decay by burning energy to maintain themselves in a \textit{nonequilibrium steady-state} (NESS) \cite{Schrodinger1944whatislife}, which has since led to the development of a range of techniques for analysing the properties of a NESS directly from observed, stochastic trajectories of a diffusion \cite{battle2016brokendetailedbalance,Gnesotto2018brokendetailedbalance,Frishman2020stochasticforce,Li2019quantifying}. Central to many of these techniques, is a \textit{discretisation} of phase-space into discrete \textit{voxels}, where the evolution of a trajectory in a continuum can be approximated as a sequence of discrete states, as shown in Fig. \ref{fig: traj coarsegrain}. These methods, and their derivatives, have been used to analyse the nonequilibrium dynamics of many biological systems \cite{battle2016brokendetailedbalance,Gnesotto2018brokendetailedbalance,lynn2021detailedbalance,nartallokaluarachchi2024decomposing, Paijmans2017kai, Kimmel2018cellstate}. Nevertheless, such approaches represent a \textit{coarse-graining}, where much of the underlying continuous information is lost. More specifically, it is known that coarse-graining can obscure the properties of a NESS in a complex, and somewhat unpredictable, manner \cite{Esposito2012coarsegraining}. For example, coarse-graining in both space and time implies that the entropy produced can only be lower bounded from observations of the coarse-grained dynamics \cite{Esposito2012coarsegraining}. Moreover, coarse-graining introduces memory effects, rendering the dynamics non-Markovian \cite{seifert2019inference}. In turn, Markovian estimates of the entropy produced are also lower-bounds on the true value \cite{Schwarz2024Memory}. Within statistical mechanics, coarse-graining has long been a powerful technique to simplify and reduce complex models through techniques like the \textit{renormalisation group} \cite{Wilson1983renormalisation}, or \textit{hydrodynamics} \cite{Marchetti2013hydrodynamics}. In fact, the stochastic formulation of Brownian motion is itself an abstraction that is obtained by coarse-graining molecular dynamics at a finer scale \cite{Schilling2022coarse}.\\
\\
Here, we shed light on the particular problem of coarse-graining a nonequilibrium diffusion with a set of discrete states. Moreover, we focus our work on the properties of the NESS in both the original diffusion and its discrete-approximation. We begin by introducing the mathematical frameworks for both continuous- and discrete-state stochastic processes and their NESS. Following this, we use a \textit{finite-volume approximation} (FVA) to derive an effective discrete-state Markov process from a coarse-grained nonequilibrium diffusion in the plane. We show, analytically, that the entropy production rate (EPR) of a continuous diffusion and its discrete-state approximation converge in the limit, which we confirm with solvable examples. Next, we use our approach to investigate NESS in unsolvable diffusions, including the stochastic van der Pol and frustrated Kuramoto oscillators. We then discuss the statistical inference of a discrete-state process from observed trajectories and show that the underlying distribution of a process can be well-approximated with a simple inference scheme. However, our results show that the EPR of an inferred process is a weak approximation of the true value. Nevertheless, we employ discrete-state models to perform `testing' of trajectories to indicate if they originate from a process in a NESS. We illustrate this statistical approach on both sampled paths and real-world trajectories from schooling fish, where we determine that the stationary dynamics underlying their movement are not out of equilibrium. Our results provide a general framework for deriving coarse-grained, discrete-state models from an underlying nonequilibrium diffusion. In particular, we show how coarse-grained models can preserve particular features of a NESS, and how such models may be inferred from observed, stochastic trajectories, which has implications for the analysis of biological and physical systems.
\section{Stochastic processes and nonequilibrium steady-states}
\label{sec: stoch proc}

We begin by introducing the mathematical frameworks for defining \textit{nonequilibrium steady-states} in both continuous- and discrete-state spaces.
\subsection{Diffusions in continuous-space: Stochastic differential equations and the Fokker-Planck equation}
\label{subsec: diffusive}

A continuous-space diffusion, $\{X(t)\in \mathbb{R}^d: t>0\}$, is defined as the solution to the \textit{stochastic differential equation} (SDE)
\begin{align}
    dX(t) = f(X(t))\;dt + \Sigma(X(t))\;dW(t), 
\end{align}
where $f:\mathbb{R}^d\rightarrow\mathbb{R}^d$ is a vector field that defines the deterministic forcing, known as the \textit{drift}, and $D(x) = \frac{1}{2}\Sigma(x)\Sigma^{\top}(x) : \mathbb{R}^{d} \rightarrow \mathbb{R}^{d\times d}$ is the \textit{diffusion matrix}, which is positive definite \cite{pavliotis2014stochproc}. {\color{red} When using spatially-dependent diffusion, we will use the Itô convention for stochastic integration \cite{pavliotis2014stochproc}.\footnote{\color{red} It is worth noting that the choice of convention for stochastic integration can lead to important differences in nonequilibrium properties like time-reversibility \cite{vanKampen1981itostratonovich,Ayala2025reversibility}.}} The \textit{density} of the process, $p(x,t)$, is defined as the probability density of the process attaining a value of $x\in\mathbb{R}^d$ at time $t>0$. Its dynamics are given by the \textit{Fokker-Planck} (FP) equation
\begin{align}
    \partial_tp(x,t) & = -\nabla\cdot J(x,t),\\
    J(x,t)& = f(x)p(x,t) - \nabla \cdot(D(x)p(x,t)),
\end{align}
where $J(x,t):\mathbb{R}^d\times \mathbb{R}\rightarrow\mathbb{R}^d$ is known as the \textit{probability flux}.\\
\\
When the process is \textit{ergodic}, it attains a unique stationary density, $\pi(x)$, which satisfies that the stationary flux is divergence-free, i.e. $\nabla \cdot \Jss = 0$, where
\begin{align}
    \Jss(x) = f(x)\pi(x)-\nabla \cdot(D(x)\pi(x)).
\end{align}
The process is said to be in a \textit{steady-state}. Such a steady-state can either be in or out of thermodynamic equilibrium. If $\Jss\equiv 0$, then the process is in an \textit{equilibrium steady-state} (ESS), otherwise, it is in a \textit{nonequilibrium steady-state} (NESS).\\
\\
Diffusions in a NESS admit the \textit{Helmholtz-Hodge decomposition} (HHD)
\begin{align}
    f &= f_{\text{rev}} + f_{\text{irr}},\\
    f_{\text{rev}}(x)& = D(x)(\nabla \log \pi(x) - \nabla \cdot D(x)),\\
    f_{\text{irr}}(x) &= \Jss(x)/\pi(x),
\end{align}
where $f_{\text{rev}}$ and $f_{\text{irr}}$ represent the \textit{time-reversible} and \textit{time-irreversible} drift respectively \cite{DaCosta_2023}. This implies that they are even and odd under time-reversal respectively.\\
\begin{figure}
    \centering
    \includegraphics[width=\linewidth]{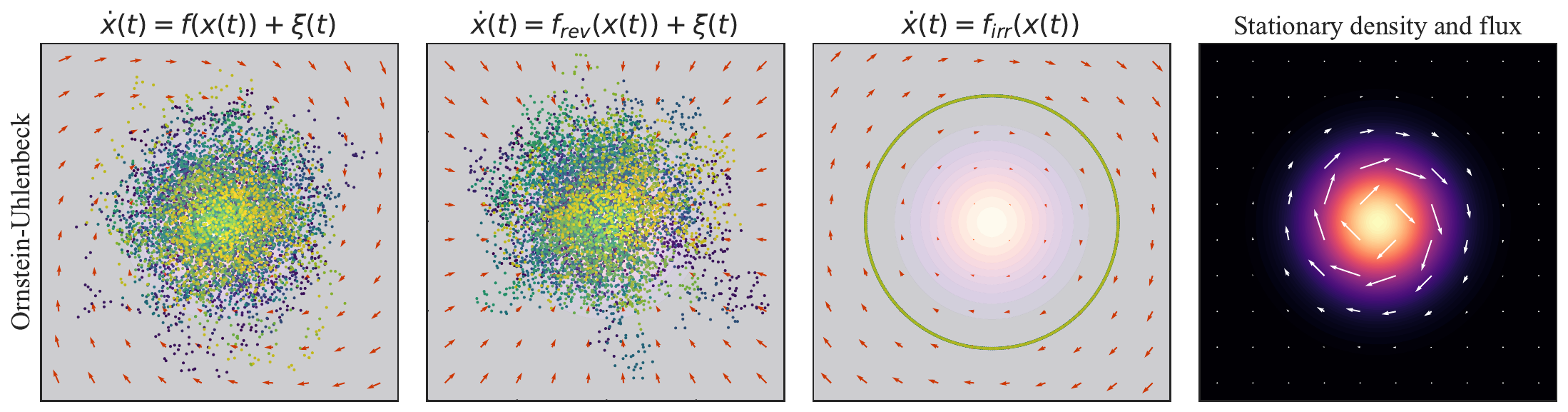}
    \caption{\textbf{Nonequilibrium steady-state}. Processes in a NESS are characterised by the presence of stationary probability flux. A process can be decomposed using the HHD into a reversible component where the drift balances the diffusive fluctuations to maintain the process at stationarity, and the irreversible component drives rotation around the stationary density.}
    \label{fig: NESS}
\end{figure}
\\
In keeping with the second law of thermodynamics, NESS thus have time-irreversible dynamics that lead to the production of entropy. The \textit{entropy production rate} (EPR) is a key quantity determining the distance of a stationary process from thermodynamic equilibrium. For a diffusion in a NESS, it is given by
\begin{align}
    \Phi & = \int_{\mathbb{R}^d}\frac{\Jss^{\top}(x)D^{-1}(x)\Jss(x)}{\pi(x)}\;dx,\label{eq: EPR FP}
\end{align}
which is only zero if the process is in a ESS, and is positive otherwise \cite{Jiang2004noneq}. Using the HHD, the EPR can be alternatively written as
\begin{align}
     \Phi & = \int_{\mathbb{R}^d}f_{\text{irr}}^{\top}(x)D^{-1}(x)f_{\text{irr}}(x)\pi(x)\;dx,\label{eq: EPR HHD}
\end{align}
which makes it clear that the EPR only depends on the irreversibility of the process, a central result in stochastic thermodynamics \cite{seifert2012thermodynamics}.

\subsection{Discrete-state processes: Continuous-time Markov chains}
\label{subsec: discrete}
Discrete-state processes are typically modelled as Markov chains, which can evolve in either continuous or discrete time-steps. We will focus on \textit{continuous-time Markov chains} (CTMC). A CTMC, $\{X(t)\in\Omega:  t>0\}$, is a stochastic process taking values in the discrete support $\Omega$ which, without loss of generality, can be assumed to be $\Omega \subseteq\mathbb{N}$. Transitions between states occur according to exponentially-distributed waiting times.\footnote{One can define a stochastic process with arbitrary, non-exponential waiting time distributions. Later on, we define such processes to be \textit{semi-Markov processes}.} The distribution of the process, $p_i(t)$, describes the probability that the process is in state $i$ at time $t$. The vector of such probabilities, $\bp(t) = (p_1(t),...)$, evolves according to the \textit{master equation} (ME)
\begin{align}
    \frac{d\bp}{dt}(t) & = L\bp,
\end{align}
where $L$ is the \textit{Laplacian} matrix \cite{Gardiner2010Methods}. The off-diagonal entries, $L_{ij}\geq 0$, represent the \textit{transition rates} from states $j$ to $i$. The diagonal entries are then defined to be $L_{ii} = -\sum_{j\neq i}L_{ji}$, which enforces that $\sum_{i} p_i(t) = 1$ as $\frac{d}{dt}(\sum_ip_i(t)) = 0$. Moreover, the waiting distribution in each state is defined to  be $\text{Exp}(-L_{ii})$ \cite{Masuda2017randomwalks}.\\
\\
If a CTMC is \textit{irreducible} and \textit{aperiodic}, then it is ergodic and will converge to a unique stationary distribution (see Ref. \cite{Ross2019probmodels} for definitions), which satisfies $L\pi = 0$. We can define a discrete counterpart to the probability flux vector field
\begin{align}
\label{eq: discrete flux}
    J_{ij} = L_{ji}\pi_i - L_{ij}\pi_j,
\end{align}
thus the steady-state is in equilibrium if and only if the flux between all pairs of states vanishes \cite{Jiang2004noneq}. This is more commonly known as the \textit{detailed balance condition}
\begin{align}
    L_{ij}\pi_j = L_{ji}\pi_i.
\end{align}
When this condition is violated, the EPR of the NESS can be computed with the Schnakenberg formula \cite{schnakenberg1976networktheory},\footnote{At first glance, the discrete and continuous forms of the EPR may not appear to have much in common. However, both are defined as the relative entropy between the forward and backward path-space measures. This can be derived both measure-theoretically, and more heuristically by taking the derivative of the Shannon entropy of either the distribution, in the case of the CTMC, or of the entropy along a stochastic trajectory, in the case of the diffusion. For further details see Refs. \cite{DaCosta_2023,Jiang2004noneq}.}
\begin{align}
    \Phi & = \frac{1}{2}\sum_{i,j}\left(L_{ij}\pi_j-L_{ji}\pi_i\right)\log\frac{L_{ij}\pi_j}{L_{ji}\pi_i}. \label{eq: schnak}
\end{align}
Similarly, the dynamics of the steady-state are reversible in an ESS and irreversible in a NESS \cite{roldan2014thesis}.
\section{Discrete-state approximations of diffusions}
\label{sec: approximation}
In this next section, we focus on developing and analysing a coarse-grained model of a nonequilibrium diffusion. We will do so by approximating it with a discrete-state Markov chain. Whilst it is important to note that such a coarse-grained model would be non-Markovian, deriving a model with arbitrary waiting-time distributions remains intractable for general multivariate diffusions. Instead, we focus on developing a Markovian approximation of the coarse-grained dynamics using a \textit{finite-volume approximation} (FVA), thus approximating the FP equation with a ME.
\subsection{Non-Markovian approaches}
\label{sec: nonmarkov}
Coarse-graining a Markovian diffusion with discrete-states leads to both memory effects and non-Markovian dynamics, in the same way that coarse-graining a Markov chain via state-aggregation induces memory in discrete-state dynamics \cite{Schwarz2024Memory,faccin2021stateaggregations}. This typically leads to both non-exponential waiting times, as well as non-Markovian transitions.\footnote{It is important to note that these are separate phenomena. Whilst non-exponential waiting times imply a semi-Markov process, coarse-graining can also induce more disruptive memory effects where transition rates between states depend on the trajectory at large.} Previous approaches have used first-passage time analysis and asymptotic expansions in an attempt to obtain the waiting-time distributions of a discrete-state approximation of prototypical, univariate diffusions \cite{Gernert2014waiting,Falasco2021localdetailed,Ghosala2022inferring,Meyberg2024entropy}. Such an analysis is closely linked to approaches for estimating the EPR from semi-Markov processes using waiting-time distributions \cite{martinez2019inferring,Skinner2021estimating}.\footnote{A \textit{semi-Markov process} is a discrete-state process with arbitrary waiting-time distributions \cite{Ross2019probmodels}.} Whilst such approaches more accurately describe the coarse-grained dynamics, they are typically limited in the complexity of the dynamics that can be applied to. Moreover, the statistical inference of semi-Markov processes from experimental data is more involved, hence data analytic methods typically use a Markovian model \cite{battle2016brokendetailedbalance,lynn2021detailedbalance,nartallokaluarachchi2024decomposing}. In discrete-time, it is simple to infer higher-order Markov models from sequence data, and thus reveal the effects of memory in real stochastic dynamics \cite{rosvall2014memory, Sahasrabuddhe2025concise}\\
\\
Other related works include the coarse-graining of a microscopic Markov process with an SDE in the continuum, or `diffusion', limit, where the convergence of the EPR has also been studied \cite{busiello2019coarsegrained}.
\subsection{A Markovian approximation}
\label{sec: markov approx}
To develop our Markovian approximation, we will consider 2-dimensional diffusions with independent sources of noise,
\begin{align}
    dx(t) &= f^x(x,y)\;dt + \sigma^{x}(x,y)\;dW^x(t),\\
    dy(t) &= f^y(x,y)\;dt + \sigma^{y}(x,y)\;dW^y(t),\notag
\end{align}
which leads to a diagonal diffusion matrix with entries
\begin{align}
    D(x,y) = \begin{pmatrix}D^{x}(x,y)& 0\\
    0 & D^y(x,y)
    \end{pmatrix}= \frac{1}{2}\begin{pmatrix}(\sigma^{x}(x,y))^2& 0\\
    0 & (\sigma^y(x,y))^2
    \end{pmatrix}.
\end{align}
This, in turn, leads to a FP equation of the form
\begin{align}
    \partial_tp(x,y,t)& = -\frac{\partial J^x(x,y,t)}{\partial x} -\frac{\partial J^y(x,y,t)}{\partial y}, \label{eq: 2d FP}
\end{align}
where the flux field is given by
\begin{align}
    J(x,y,t) &= \begin{pmatrix}
        J^x(x,y,t)\\
        J^y(x,y,t)
    \end{pmatrix},\\
    &=\begin{pmatrix}
        f^x(x,y)- \partial_xD^x(x,y)\\
        f^y(x,y)- \partial_yD^y(x,y)
    \end{pmatrix}p(x,y,t) - \begin{pmatrix}
        D^x(x,y)\partial_xp(x,y,t)\\
        D^y(x,y)\partial_yp(x,y,t)
    \end{pmatrix}.\notag
\end{align}
\begin{figure}
    \centering
    \includegraphics[width=0.85\linewidth]{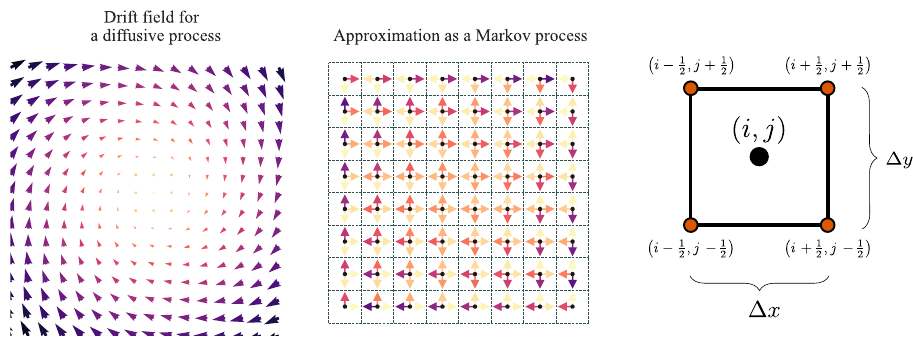}
    \caption{\textbf{Approximating a diffusion as a Markov process}. We aim to derive an approximation of a continuous diffusion as a discrete-state Markov chain using a finite-volume approximation. We use a rectangular grid to coarse-grain $\mathbb{R}^2$ into a set of \textit{volumes}, where we approximate the flux across the boundary, resulting in a CTMC.}
    \label{fig: markov approx}
\end{figure}
First, we discretise $\mathbb{R}^2$ into a grid of rectangular cells of dimension $\Delta x \times \Delta y$. We denote the midpoint of each cell to be $(x_i,y_j)$, thus its boundary is the rectangle $R_{ij} = [x_{i-\frac{1}{2}},x_{i+\frac{1}{2}}]\times[y_{j-\frac{1}{2}},y_{j+\frac{1}{2}}]$, as shown in Fig. \ref{fig: markov approx}. As described in Sec. \ref{sec: intro}, the discretisation of a trajectory is performed by assigning a discrete-state to each voxel in the discretised space. In this case, the process is in state $k=(i,j)$ at time $t$ if the underlying diffusion is in the corresponding cell i.e. $(x(t),y(t))\in B_{ij}$. The corresponding ME describes the evolution of $p_{i,j}(t)$, the probability that the process is in state $(i,j)$.\\
\\
A FVA of Eq. (\ref{eq: 2d FP}) yields
\begin{align}
    \frac{dp_{i,j}}{dt}& = -\frac{J^x_{i+\frac{1}{2},j}-J^x_{i-\frac{1}{2},j}}{\Delta x} - \frac{J^y_{i,j+\frac{1}{2}}-J^y_{i,j-\frac{1}{2}}}{\Delta y}, \label{eq: discrete FP}
\end{align}
where the subscript $k,l$ implies the function is evaluated at $(x_k,y_l)$. The expressions on the right of Eq. (\ref{eq: discrete FP}) represent the discrete flux along the bounding edges of the rectangle. In practice, we assume zero-flux at the boundary, which is chosen sufficiently far from the centre of mass of the stationary density, to simulate an infinite domain.\\
\\
It remains to derive expressions for the numerical flux which ultimately leads to a ME of the form
\begin{align}
    \frac{dp_{i,j}}{dt}& = \sum_{(k,l)} L_{(i,j),(k,l)}p_{k,l}.
\end{align}
A numerical scheme yields a valid ME if $L_{(i,j),(k,l)}>0$ for all off-diagonal entries ($(k,l)\neq (i,j)$), and probability is conserved i.e. the columns of $L$ sum to 0. In Ref. \cite{strang2020applications}, Strang considered the centred-difference approximation for the FP equation in 1D, which only yields a valid ME under step-size conditions. {\color{red} As a result, the problem of deriving a valid approximation is closely related to that of developing a structure-preserving scheme for the FP equation \cite{Holubec2019physically,Heida2021finitevolume}. We will adopt the \textit{Scharfetter-Gummel} discretisation \cite{Frensley2004scharfettergummel}, which alleviates step-size conditions.}

\subsection{The Scharfetter-Gummel discretisation}
The \textit{Scharfetter-Gummel} (SG) discretisation is a structure-preserving scheme for the FP equation that yields the following approximation for the flux across the boundary,
\begin{align}
\label{eq: sg}
    J^x_{i+\frac{1}{2},j}&=\frac{f^x_{i+\frac{1}{2},j}\left(p_{i,j}-e^{-f^x_{i+\frac{1}{2},j}\Delta x/D^x_{i+\frac{1}{2},j}}p_{i+1,j}\right)}{1-e^{-f^x_{i+\frac{1}{2},j}\Delta x/D^x_{i+\frac{1}{2},j}}},
\end{align}
with analogous expressions for the remaining flux terms \cite{Frensley2004scharfettergummel} (for a derivation, see App. \ref{app: SG disc}). This naturally leads to a ME of the form
\begin{align}
    \frac{dp_{i,j}}{dt}& = L_{(i,j)(i,j)} + \sum_{k=i\pm1}L_{(i,j)(k,j)}+\sum_{k=j\pm1}L_{(i,j)(i,k)},
\end{align}
with off-diagonal transition rates given by
\begin{align}
    L_{(i,j),(i+1,j)}& = \frac{f^x_{i+\frac{1}{2},j}e^{-f^x_{i+\frac{1}{2},j}\Delta x/D^x_{i+\frac{1}{2},j}}}{\Delta x\left(1-e^{-f^x_{i+\frac{1}{2},j}\Delta x/D^x_{i+\frac{1}{2},j}}\right)},\\
    L_{(i,j),(i-1,j)}& = \frac{f^x_{i-\frac{1}{2},j}}{\Delta x\left(1-e^{-f^x_{i-\frac{1}{2},j}\Delta x/D^x_{i-\frac{1}{2},j}}\right)},\\
    L_{(i,j),(i,j+1)}& = \frac{f^y_{i,j+\frac{1}{2}}e^{-f^y_{i,j+\frac{1}{2}}\Delta y/D^y_{i,j+\frac{1}{2}}}}{\Delta y\left(1-e^{-f^y_{i,j+\frac{1}{2}}\Delta y/D^y_{i,j+\frac{1}{2}}}\right)},\\
    L_{(i,j),(i,j-1)}& = \frac{f^y_{i,j-\frac{1}{2}}}{\Delta y\left(1-e^{-f^y_{i,j-\frac{1}{2}}\Delta y/D^y_{i,j-\frac{1}{2}}}\right)},
\end{align}
which are necessarily positive.\footnote{In cases, where the denominator vanishes, which occurs when the drift vanishes, then we can re-derive the SG discretisation to obtain $L_{(i,j),(i+1,j)} = D^{x}_{i+\frac{1}{2},j}/\Delta x^2$, and analogous terms for the other transitions \cite{Frensley2004scharfettergummel}.} Moreover, the diagonal rates are given by
\begin{align}
    L_{(i,j),(i,j)}= &-\frac{f^x_{i+\frac{1}{2},j}}{\Delta x\left(1-e^{-f^x_{i+\frac{1}{2},j}\Delta x/D^x_{i+\frac{1}{2},j}}\right)} - \frac{f^x_{i-\frac{1}{2},j}e^{-f^x_{i-\frac{1}{2},j}\Delta x/D^x_{i-\frac{1}{2},j}}}{\Delta x\left(1-e^{-f^x_{i-\frac{1}{2},j}\Delta x/D^x_{i-\frac{1}{2},j}}\right)}\\
    &- \frac{f^y_{i,j+\frac{1}{2}}}{\Delta y\left(1-e^{-f^y_{i,j+\frac{1}{2}}\Delta y/D^y_{i,j+\frac{1}{2}}}\right)}-\frac{f^y_{i,j-\frac{1}{2}}e^{-f^y_{i,j-\frac{1}{2}}\Delta y/D^y_{i,j-\frac{1}{2}}}}{\Delta y\left(1-e^{-f^y_{i,j-\frac{1}{2}}\Delta y/D^y_{i,j-\frac{1}{2}}}\right)}.\notag
\end{align}
It is simple to check that the sum
\begin{align}
    L_{(i+1,j),(i,j)}+ L_{(i-1,j),(i,j)}+L_{(i,j+1),(i,j)}+L_{(i,j-1),(i,j)}+L_{(i,j),(i,j)}&= 0,
\end{align}
by replacing $i$ by $i\pm1$ and $j$ by $j\pm1$ into the derived rates. As a result, this scheme yields a valid ME as it has non-negative rates between states and conserves probability.\\
\\
It is well-known that FVA conserve probability \cite{LeVeque2007finitedifference}.\footnote{i.e. the numerical solution at every time integrates to 1. This is equivalent to the derived Laplacian having columns that sum to 0.} Moreover, the SG discretisation preserves the stationary distribution and many thermodynamic properties of the continuous solution \cite{Schlichting2022Scharfetter}, hence it is well-suited to this application.
\subsubsection{Entropy production in the Scharfetter-Gummel discretisation}
\label{sec: convergence}

In this section, we will show that the EPR of the discrete approximation converges to that of the underlying diffusion process. We begin by assuming that the discrete process is stationary with distribution $\pi_{i,j}= \Delta x \Delta y \,\pi(x_i,y_j) + O(\Delta x^3\Delta y+\Delta x\Delta y^3)$, which is a midpoint approximation of the true stationary density $\pi(x,y)$. As each cell only receives transitions from the four adjacent cells, the EPR of Eq. (\ref{eq: schnak}) reduces to
\begin{align}
    \Phi = & \frac{1}{2}\sum_{i,j}\Phi_{(i,j),(i+1,j)}+ \Phi_{(i,j),(i-1,j)} + \Phi_{(i,j),(i,j+1)} + \Phi_{(i,j),(i,j-1)},
\end{align}
where each contribution is defined by
\begin{align}
    \Phi_{(i,j)(i+1,j)} & = \left(L_{(i,j),(i+1,j)}\pi_{i+1,j} - L_{(i+1,j),(i,j)}\pi_{i,j}\right) \log \frac{L_{(i,j),(i+1,j)}\pi_{i+1,j}}{L_{(i+1,j),(i,j)}\pi_{i,j}},
\end{align}
with analogous expressions for the remaining terms. Substituting in the expressions derived from the SG discretisation, we obtain
\begin{align}
    \Phi_{(i,j)(i+1,j)} & = \frac{f^x_{i+\frac{1}{2},j}}{\Delta x\left(1-e^{-f^x_{i+\frac{1}{2},j}\Delta x/D^x_{i+\frac{1}{2},j}}\right)}\left(e^{-f^x_{i+\frac{1}{2},j}\Delta x/D^x_{i+\frac{1}{2},j}}\pi_{i+1,j}-\pi_{i,j}\right)\left(-\frac{f^x_{i+\frac{1}{2},j}\Delta x}{D^x_{i+\frac{1}{2},j}} + \log\frac{\pi_{i+1,j}}{\pi_{i,j}}\right).
\end{align}
Next, we perform an expansion to derive the leading order entropy production along an edge,
\begin{align}
    \Phi_{(i,j)(i+1,j)} &= \Delta x\Delta y \frac{\left(-f^x_{i+\frac{1}{2},j}\pi(x_i,y_j) + D^x_{i+\frac{1}{2},j}[\partial_x\pi(x,y)]_{i,j}\right)^2}{D^x_{i+\frac{1}{2},j}\pi(x_i,y_j)} + O(\Delta x^2\Delta y + \Delta y^2 \Delta x),
\end{align}
where we are using that the stationary distribution is convergent. Pairing the edges in the $x$-direction, we have that
\begin{align}
    \frac{1}{2}\left(\Phi_{(i,j)(i+1,j)} + \Phi_{(i,j)(i-1,j)}\right)& = \Delta x\Delta y\frac{(J^x_{i,j})^2}{D^x_{i,j}\pi(x_i,y_j)} +O(\Delta x^2\Delta y + \Delta y^2 \Delta x), 
\end{align}
where $J$ is the probability flux associated with the stationary density $\pi$. Combining the edges in the $y$-direction, we have that the EPR is,
\begin{align}
    \Phi & = \Delta x \Delta y\sum_{(i,j)} \frac{(J^x_{i,j})^2}{D^x_{i,j}\pi(x_i,y_j)} + \frac{(J^y_{i,j})^2}{D^y_{i,j}\pi(x_i,y_j)} + O(\Delta x+\Delta y).
\end{align}
Finally, we can take the limit as $\Delta x, \Delta y$ go to zero, 
\begin{align}
    \lim_{\Delta x, \Delta y \rightarrow0} \Phi & = \int\int \frac{(J^x(x,y))^2}{D^x(x,y)\pi(x,y)}+\frac{(J^y(x,y))^2}{D^y(x,y)\pi(x,y)} \,dx\,dy,
\end{align}
which is precisely Eq. (\ref{eq: EPR FP}) in the case of a 2D process with diagonal diffusion matrix i.e. the EPR of the CTMC converges to that of the underlying diffusion process.

{\color{red} 

\subsubsection{Recovering the drift and diffusion coefficients} 

Given the Laplacian matrix, $L$, we can  recover the drift and diffusion at discrete points in the grid. To show this explicitly, we consider the rates
\begin{align}
    L_{(i,j),(i+1,j)}& = \frac{f^x_{i+\frac{1}{2},j}e^{-\xi}}{\Delta x (1-e^{-\xi})}, && L_{(i+1,j),(i,j)}=\frac{f^x_{i+\frac{1}{2},j}}{\Delta x(1-e^{-\xi})},
\end{align}
where $\xi = f^x_{i+\frac{1}{2},j}\Delta x/D^x_{i+\frac{1}{2},j}$. These are the transition rates from $(i+1,j)$ to $(i,j)$ and its reverse, respectively. We momentarily drop the subscript for convenience, and take the difference in the rates to obtain,
\begin{align}
    L_{(i+1,j),(i,j)} - L_{(i,j),(i+1,j)}& = \frac{f^x - f^xe^{-\xi}}{\Delta x (1-e^{-\xi})}=\frac{f^x}{\Delta x},
\end{align}
thus we  recover the drift with
\begin{align}
    f^x_{i+\frac{1}{2},j} & = \Delta x\left(L_{(i+1,j)(i,j)} - L_{(i,j)(i+1,j)}\right).
\end{align}
The remaining drift terms can be obtained analogously. To recover the diffusion, we take the log-ratio of the rates to obtain
\begin{align}
    \xi = \log\left(\frac{L_{(i+1,j)(i,j)}}{L_{(i,j)(i+1,j)}}\right) = \frac{f^x\Delta x}{D^x}.
\end{align}
We then use our previous expression for the drift to write the diffusion as
\begin{align}
    D^x_{i+\frac{1}{2},j}&=\frac{\Delta x^2\left(L_{(i+1,j)(i,j)} - L_{(i,j)(i+1,j)}\right)}{\log\left(L_{(i+1,j)(i,j)}/L_{(i,j)(i+1,j)}\right)}.
\end{align}
The remaining diffusion coefficients can be recovered in a similar way.

\begin{remark}[A variational perspective on convergence from Markov chains to diffusions]
\label{remark: variational}
It is not immediately obvious that the NESS properties of a CTMC converge to that of the SDE, even when the CTMC is derived from the drift and diffusion of the SDE. In the former, fluctuations have Poissonian statistics stemming from the exponentially-distributed waiting times between transitions, whilst in the latter the fluctuations are driven by Gaussian noise. In order to develop further intuition on this convergence, we consider the \textit{variational approach} to characterising a NESS, which combines macroscopic fluctuation theory \cite{Bertini2015macroscopic}, and the large deviation principle \cite{Touchette2013largedeviation}.\\
\\
It is useful to define the \textit{dissipation potential} and its complex conjugate $(\Psi, \Psi^*)$, which capture the thermodynamic cost of observing a given fluctuation in the flux $J$. Moreover, at steady-state the EPR is given by $\Phi = \Psi + \Psi^*$ \cite{mielke2014gradientflows, Mielke2016generalisation, Mielke2023saddle, Kaiser2018canonical}. It is well-known that the dissipation potential for a diffusive process is quadratic in the flux, leading to the characteristic quadratic form for the EPR, as in Eq. (\ref{eq: EPR FP}) and (\ref{eq: EPR HHD}). This implies a `linear response', which, in turn, arises from the Gaussian nature of the fluctuations. On the other hand, the dissipation potential of a CTMC is non-quadratic \cite{mielke2014gradientflows,Kaiser2018canonical}. However, as discussed and illustrated in App. \ref{app: variational}, in the continuum limit the potential, and therefore the EPR, is approximately quadratic in the flux. This argument provides a further intuition for the convergence in Sec. \ref{sec: convergence}.
\end{remark}
}

\section{Example processes and numerical experiments}
In this section, we consider our scheme applied to both solvable and unsolvable nonequilibrium diffusions. In the first instance, we confirm with numerical experiments that the stationary distribution and EPR of the discrete process converges for the \textit{Ornstein-Uhlenbeck} process and stochastic \textit{Hopf oscillator}, which have solvable steady-states. Next, we apply our method to investigate the stationary distribution for the stochastic \textit{van der Pol} oscillator and a pair of coupled, frustrated, stochastic \textit{Kuramoto} oscillators, for which there is no analytical form for the steady-state.
\subsection{Solvable models}
\subsubsection{The Ornstein-Uhlenbeck process}
The \textit{Ornstein-Uhlenbeck} (OU) process is a multivariate diffusion process that models interacting degrees of freedom under both friction and fluctuations \cite{Godreche2018OU}. It is given by the SDE
\begin{align}
        dX(t) &= -BX(t) dt + \Sigma \; dW(t),
\end{align}
where $B\in \mathbb{R}^d \times \mathbb{R}^d$ is the \textit{friction matrix} and $D=\frac{1}{2}\Sigma\Sigma^{\top}$ is the \textit{diffusion matrix}. If all eigenvalues of $B$ have positive real-parts, the process converges to a zero-mean Gaussian stationary density given by\footnote{As we denote the stationary density as $\pi$, we use $\bm{\pi}$ to denote the mathematical constant.}
\begin{align}
        \pi(x) & = (2\bm{\pi})^{-d/2}(\det S)^{1/2}\exp \left( -\frac{1}{2}x^{\top}S^{-1}x \right),
\end{align}
which corresponds to the multivariate normal distribution $\mathcal{N}_d(0,S)$, where $S$ is the covariance matrix satisfying the Lyapunov equation
\begin{align}
\label{eq: lyapunov}
     BS+SB^{\top} = 2D.
\end{align}
In general, steady-states of the OU are out of equilibrium. More specifically, the steady-state is in equilibrium if and only if $BD=DB^{\top}$ which implies $S=B^{-1}D$ \cite{lax1960fluctuationsnonequilibrium}. Otherwise, the stationary probability flux is given by
\begin{align}
    \Jss(x) = \mu x\pi(x),
\end{align}
where $\mu = DS^{-1}-B$, which then allows for an explicit calculation of the EPR. We first define $Q = BS-D$, which implies that $\mu = - QS^{-1}$ and $Q=-Q^{\top}$ from the Lyapunov equation, Eq. (\ref{eq: lyapunov}). {\color{red}With Eq. (\ref{eq: EPR FP}) and basic results for Gaussian variables, we can obtain $\Phi = \Tr(B^{\top}D^{-1}Q)$ \cite{Godreche2018OU,DaCosta_2023}}.
\\\\
In particular, we will consider the 2D system,
\begin{align}
\label{eq: OU ex}
    d\begin{pmatrix}
        x(t)\\
        y(t)
    \end{pmatrix}& = -\begin{pmatrix}
        2 & -\theta\\
        \theta & 2
    \end{pmatrix}\begin{pmatrix}
        x\\
        y
    \end{pmatrix}\;dt + \begin{pmatrix}
        \sigma & 0\\
        0 & \sigma
    \end{pmatrix}dW(t),
\end{align}
which has covariance $S= \frac{\sigma^2}{4}I$ and $\Phi = \theta^2$.
\subsubsection{The stochastic Hopf oscillator}
The \textit{Hopf oscillator} (HO) is a nonlinear process given by the equations\footnote{Also known as the \textit{Stuart-Landau oscillator} \cite{Matthews1990phasediagram}.}
\begin{align}
    dx(t)&= (a-x^2-y^2)x-\omega y\; dt+ \sigma dW_x(t),\\
    dy(t)&= (a-x^2-y^2)y+\omega x\; dt+ \sigma dW_y(t),\notag
\end{align}
where $\omega$ is the natural frequency and $\sigma$ is the noise intensity. In the absence of noise, for $a<0$ the system decays to the origin which is an attracting fixed point. At $a=0$, the system goes through a Hopf bifurcation, resulting in limit-cycle oscillations for $a>0$ \cite{strogatz2008nonlinear}. This system has become a useful model of neural dynamics \cite{nartallokalu2025review}, and hair-cell bundles \cite{Sheth2018hair}.
\\
\\
To study the NESS of this process in the presence of additive noise, we consider the FP equation in the alternative form
\begin{align}
    \partial_tp & = -(\nabla\cdot f)p -(f\cdot \nabla p)+\frac{\sigma^2}{2}\nabla^2p,
\end{align}
and replace $\nabla, \nabla \cdot,$ and $\nabla^2$ with their expressions in polar coordinates i.e. $x=r\cos \theta$ and $y=r\sin\theta$. We obtain
\begin{align}
    \partial_tp(r,\theta,t)&= \left(4r^2-2a\right)p-\left(\left(a-r^2\right)r\frac{\partial p}{\partial r} + \frac{\omega }{r}\frac{\partial p}{\partial \theta}\right) + \frac{\sigma^2}{2}\left(\frac{\partial^2p}{\partial r^2} + \frac{1}{r}\frac{\partial p }{\partial r} + \frac{1}{r^2}\frac{\partial^2p}{\partial \theta ^2}\right).
\end{align}
In order to solve for the steady-state, we use that the limit-cycle behaviour of the deterministic system is radially symmetric, and thus assume that the stationary density is also i.e. $\frac{\partial p}{\partial \theta } = \frac{\partial^2 p}{\partial \theta}=0$, which implies the distribution takes the form
\begin{align}
    \pi(r,\theta) = \frac{1}{2\bm{\pi}}\pi(r),
\end{align}
where $\pi(r)$ satisfies
\begin{align}
\label{eq: second order ode hopf}
0 & = \left(4r^2-2a\right)\pi+ \left(\frac{\sigma^2}{2r}+r^3-ar\right)\frac{d\pi}{d r} + \frac{\sigma^2}{2}\frac{d^2 \pi}{d r^2},
\end{align}
which is solved by a \textit{Boltzmann-Gibbs} density\footnote{It is not immediately obvious why a solution of the form of Eq. (\ref{eq: stationary hopf}) solves Eq. (\ref{eq: second order ode hopf}). However, it is intuitive that the irreversible, oscillatory circulation is driven by the $\theta$ dynamics, which are absent here. This suggests that the dynamics in $r$ may be the negative-gradient of a scalar potential. We can see that $f^r = -\frac{dU}{dr}$ for a potential $U = \frac{1}{2}r^2(a-\frac{1}{2}r^2)$. All that remains is to check that $\pi(r)= \exp\left(-2U/\sigma^2\right)$ solves Eq. (\ref{eq: second order ode hopf}). Whilst the stationary density is a Boltzmann-Gibbs density, the steady-state is in nonequilibrium.}
\begin{align}
\label{eq: stationary hopf}
    \pi(r,\theta) = \frac{1}{Z}\exp\left(\frac{r^2}{\sigma^2}\left(a-\frac{r^2}{2}\right)\right),
\end{align}
where $Z$ is the normalising factor\footnote{An alternative approach to solving for the stationary density can be found for the case $a=\omega=1$ in Ref. \cite{Yuan2017decomposition}.}
\begin{align}
    Z & = \frac{1}{\sqrt{2}}\exp\left(\frac{a^2}{2\sigma^2}\right)\bm{\pi}^{3/2}\sigma\left(1+\text{erf}\left(\frac{a}{\sqrt{2} \sigma }\right)\right).
\end{align}
Given the stationary density, we can perform the HHD to obtain
\begin{align}
    f_{\text{rev}}(x,y)&= \begin{pmatrix}
            x(a-x^2-y^2) \\
            y(a-x^2-y^2)
            \end{pmatrix},\\
    f_{\text{irr}}(x,y)&= \begin{pmatrix}
            -\omega y \\
            \omega x
            \end{pmatrix},
\end{align}
where the irreversible component is driving circular rotation around the stationary density. Moreover, the irreversibility of the process is driven by the frequency of oscillation, and is in equilibrium only when $\omega = 0$. To confirm this, we can use Eq. (\ref{eq: EPR HHD}), to calculate the EPR to be,
\begin{align}
\label{eq: epr hopf}
    \Phi & = \frac{\bm{\pi}\omega^2}{\sigma Z}\left(2\sigma + \sqrt{2\bm{\pi}}a e^{\frac{a^2}{4 \sigma ^2}}\left(1 + \erf{\left(\frac{a}{\sqrt{2}\sigma }\right)}\right)\right),
\end{align}
which scales quadratically in $\omega$. Fig. \ref{fig: hopf ness} shows examples of the Hopf oscillator in a NESS for $a=-1$ and $a=2$. Moreover, it shows the EPR as a function of $\omega$ and $a$ for a fixed value of $\sigma$.
\subsubsection{Numerical experiments}
\begin{figure}
    \centering
    \includegraphics[width=\linewidth]{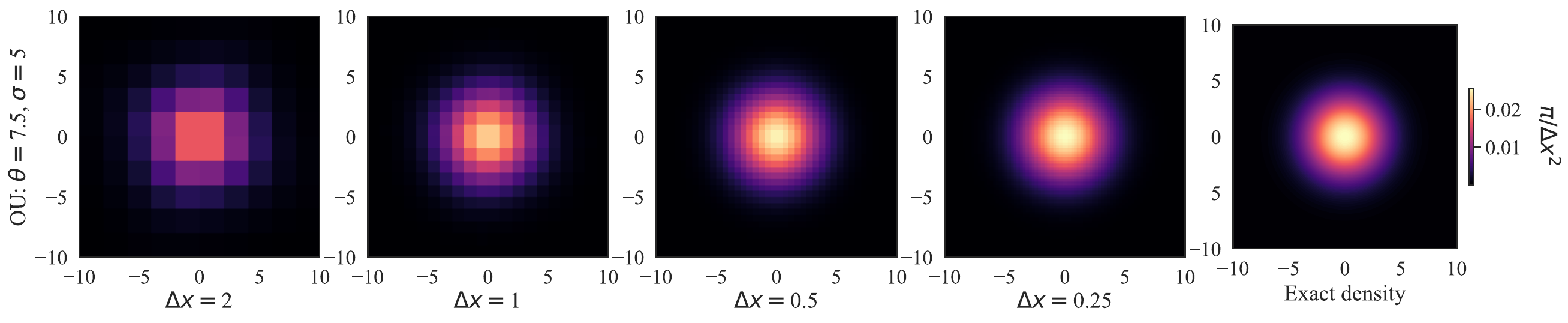}
    \caption{\textbf{Discrete-approximation of NESS in the OU.} For a fixed value of $\theta =5$, $\sigma = 5$, and a range of discretisation step-sizes, we calculate the stationary distribution of a coarse-grained OU process using the SG discretisation. Even for large step-sizes, the symmetry of the stationary density is preserved, converging to the true density as the step-size decreases.}
    \label{fig: OU conv}
\end{figure}

\begin{figure}
    \centering
\includegraphics[width=0.8\linewidth]{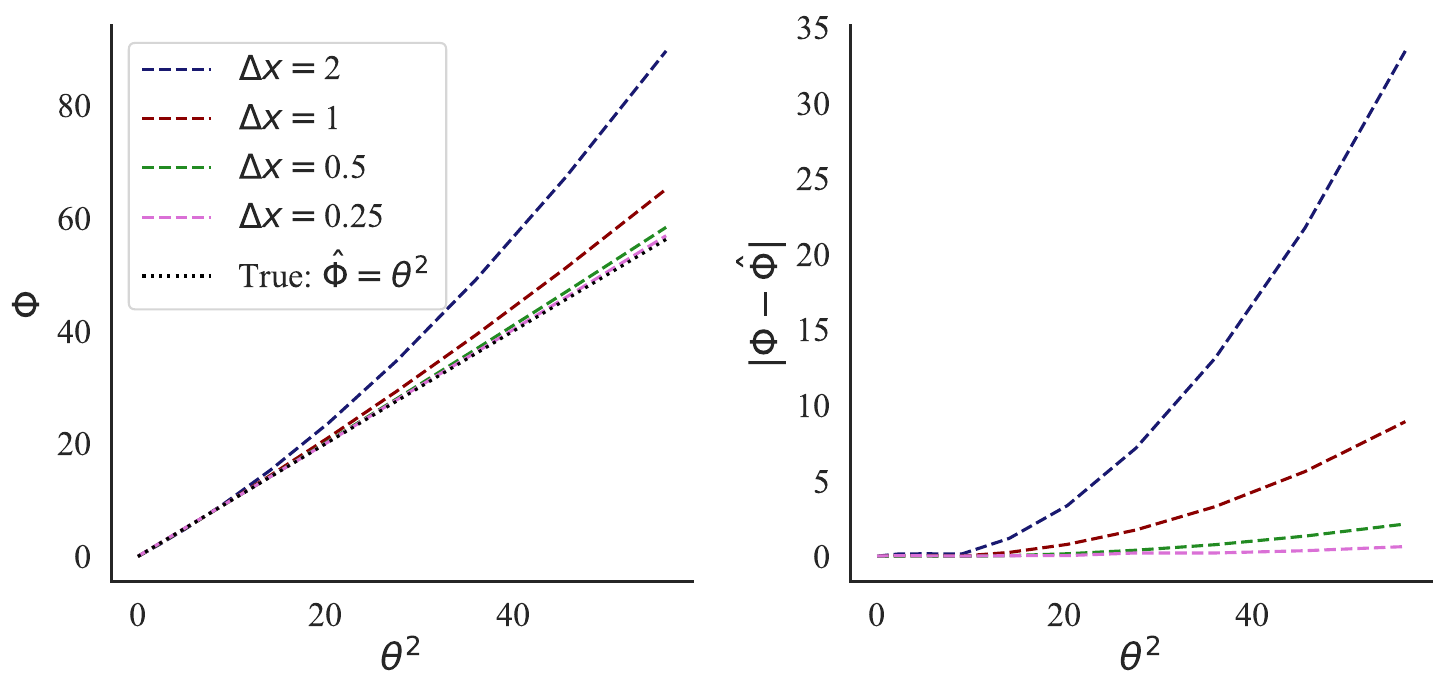}
    \caption{\textbf{EPR in a coarse-grained OU process}. We see that the EPR of the discrete-state approximation also increases approximately quadratically with $\theta$, as in the exact solution. Moreover, we can see that the EPR converges as the step-size goes to 0.}
    \label{fig: OU EPR}
\end{figure}

\begin{figure}
    \centering
    \includegraphics[width=\linewidth]{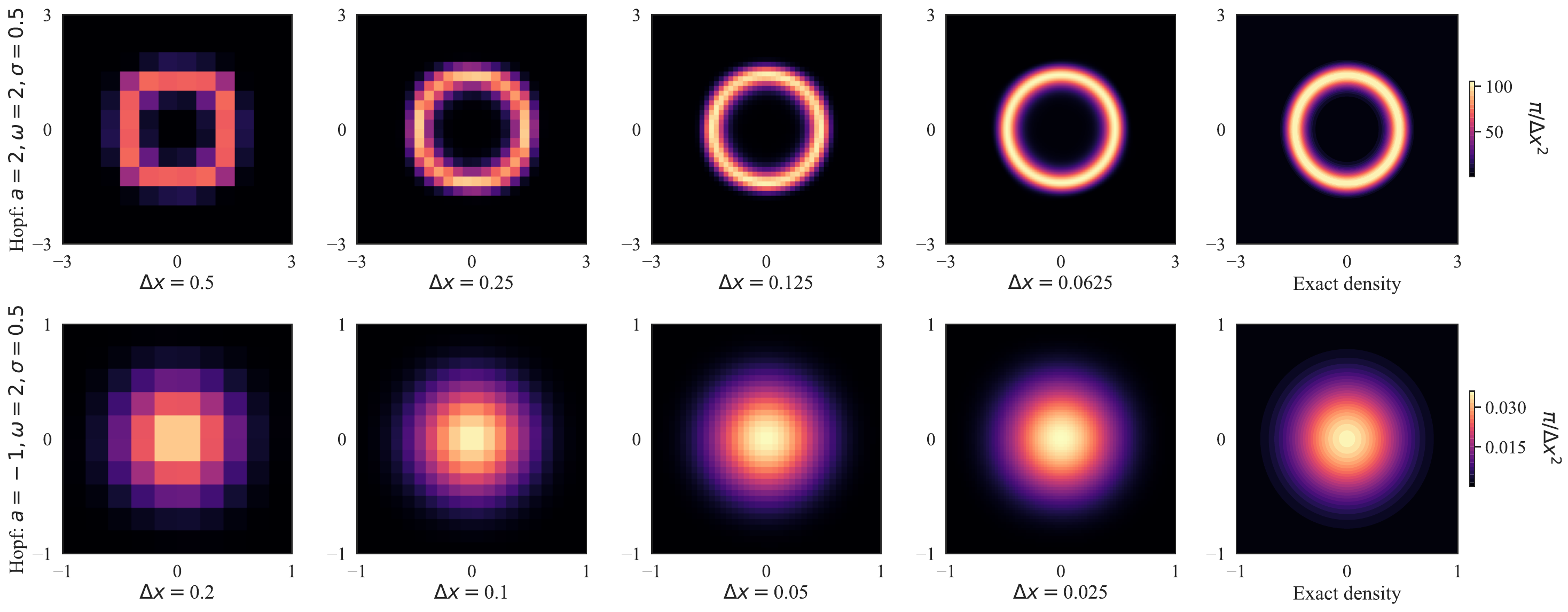}
    \caption{\textbf{Discrete-approximation of NESS in the Hopf oscillator.} For fixed values of $a\in \{-1,2\}$, $ \omega = 2$, $\sigma = 0.5$, and a range of discretisation step-sizes, we calculate the stationary distribution of a coarse-grained Hopf oscillator using the SG discretisation. Even for large step-sizes, the symmetry of the stationary density is preserved, converging to the true density as the step-size decreases.}
    \label{fig: Hopf conv}
\end{figure}

\begin{figure}
    \centering
    \includegraphics[width=\linewidth]{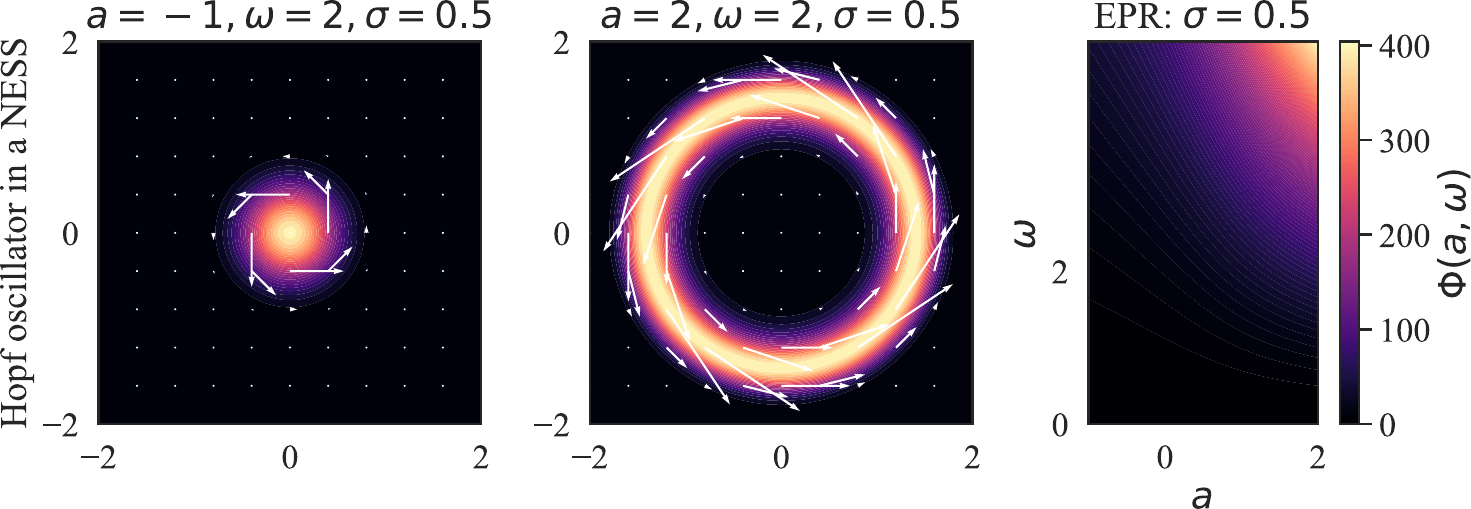}
    \caption{\textbf{Hopf oscillator in a nonequilibrium steady-state}. The stationary density and flux show that the Hopf oscillator converges to a NESS. For $a<0$, this is a distribution peaked at the origin. For $a>0$, this is a `Mexican-hat' distribution. Both show rotational probability flux due to the oscillatory dynamics. The EPR varies as a function of $\omega$ and $a$. Both $\omega$ and $a$ drive irreversible dynamics, whilst $\sigma$ drives reversible diffusion.}
    \label{fig: hopf ness}
\end{figure}

\begin{figure}
    \centering
    \includegraphics[width=\linewidth]{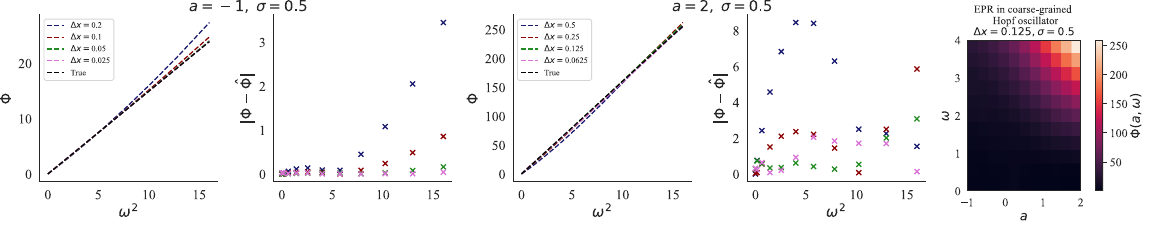}
    \caption{\textbf{EPR in the coarse-grained Hopf oscillator.} For fixed values of $a$ and $\sigma$, we can increase $\omega$ to increase the EPR of the Hopf oscillator. We show that the EPR of the underlying diffusion is closely approximated by the coarse-grained process. For $a=-1$, we see that the EPR converges as $\Delta x$ goes to 0. For $a=2$, we use larger step-sizes, and appear to be outside the asymptotic regime. Moreover, we note that for some parameters, calculating the steady-state distribution numerically can induce another source of error not accounted for in the analysis (see App. \ref{app : solving stationary}). Moreover, for a fixed step-size $\Delta x$, we show that the EPR shows the same qualitative behaviour as the true EPR of the underlying diffusion, shown in Fig. \ref{fig: hopf ness}.}
    \label{fig: Hopf EPR}
\end{figure}

We begin our numerical experiments by calculating the stationary distribution for the OU process defined in Eq. (\ref{eq: OU ex}) for fixed values of $\theta =5$ and $\sigma = 5$, whilst varying the discretisation size $\Delta x = \Delta y$ in a finite domain.\footnote{In this case, we use the domain $[-10,10]^2$ which we regard as sufficiently far from the centre of mass of the distribution that the no-flux boundaries do not significantly alter the behaviour of the discrete process. For a discussion of boundary conditions and their implementation see App. \ref{app : boundary}.} In Fig. \ref{fig: OU conv}, we see that the stationary distribution of the coarse-grained dynamics preserves the symmetry of the underlying Gaussian distribution, even for larger step-sizes. As the step-size decreases, the distribution converges to the exact density.\footnote{To compare between a discrete distribution that must \textit{sum} to 1, and a continuous density that must \textit{integrate} to 1, we normalise the discrete distribution $\pi$ by the area of each volume $\Delta x^2$ i.e. the density in cell $i$ can be approximated discretely as $\pi_i/\Delta x^2$.}\\
\\
Next, by varying the parameter $\theta$, we can vary the EPR, $\Phi = \theta^2$, of the underlying OU process. Fig. \ref{fig: OU EPR} shows the EPR of the coarse-grained process for a range of $\Delta x$ values as a function of $\theta$. Whilst all the coarse-grained dynamics follow the correct trajectory, increasing the EPR with $\theta$, the smaller step-sizes more closely approximate the true EPR. Moreover, we notice that the EPR of the coarse-grained dynamics may exceed the true EPR. Unlike the EPR of a diffusion observed at a coarse-grained level, the Markov process derived here is not subject to the inequality proved by Esposito in Ref. \cite{Esposito2012coarsegraining}, where it would necessarily lower-bound the true EPR.\\
\\
We then turn to the Hopf oscillator, where we perform similar experiments. In Fig. \ref{fig: Hopf conv}, we consider the stationary distribution of the coarse-grained Hopf oscillator for fixed values of $a=2$ (top row) and $a=-1$ (bottom row), across a range of step-sizes. As before, the coarse-grained process preserves the symmetry and shape of the stationary distribution even for larger step-sizes, for both the peaked and Mexican hat distributions, converging in the limit.\\
\\
As shown in Eq. (\ref{eq: epr hopf}), the EPR of the Hopf oscillator increases quadratically in $\omega$. In Fig. \ref{fig: Hopf EPR}, for fixed values of $a=2,-1$ and $\sigma = 0.5$, we show that the EPR of the coarse-grained process closely approximates that of the underlying diffusion, converging as the step-size decreases, as expected. Moreover, for a fixed step-size of $\Delta x = 0.125$, we explore the EPR of the coarse-grained process as a function of $a$ and $\omega$, showing the same qualitative behaviour as the exact solution in Fig. \ref{fig: hopf ness}.
\subsection{Unsolvable models}
\subsubsection{The stochastic van der Pol oscillator}
\begin{figure}
    \centering
    \includegraphics[width=\linewidth]{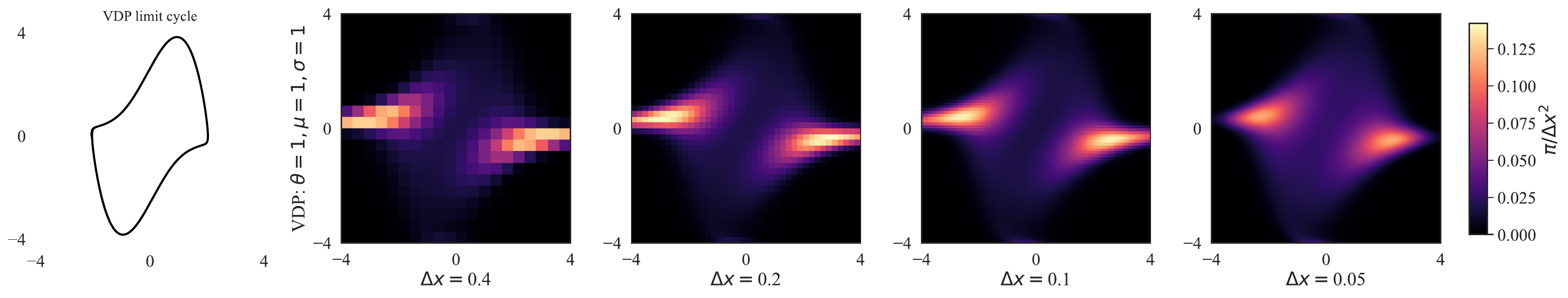}
    \caption{\textbf{Discrete-approximation of NESS in the VDP.} For a fixed value of $\theta =1$, $\mu=2$ and $\sigma=1$, we calculate the stationary distribution of a coarse-grained VDP oscillator using the SG discretisation. Even for small step-sizes, the limit cycle behaviour is present. Moreover, unlike the Hopf oscillator, the rotation is not of uniform speed leading to a build up of probability at the apexes. We consider the domain $[-4,4]^2$.}
    \label{fig: VDPStationary}
\end{figure}

\begin{figure}
    \centering
    \includegraphics[width=\linewidth]{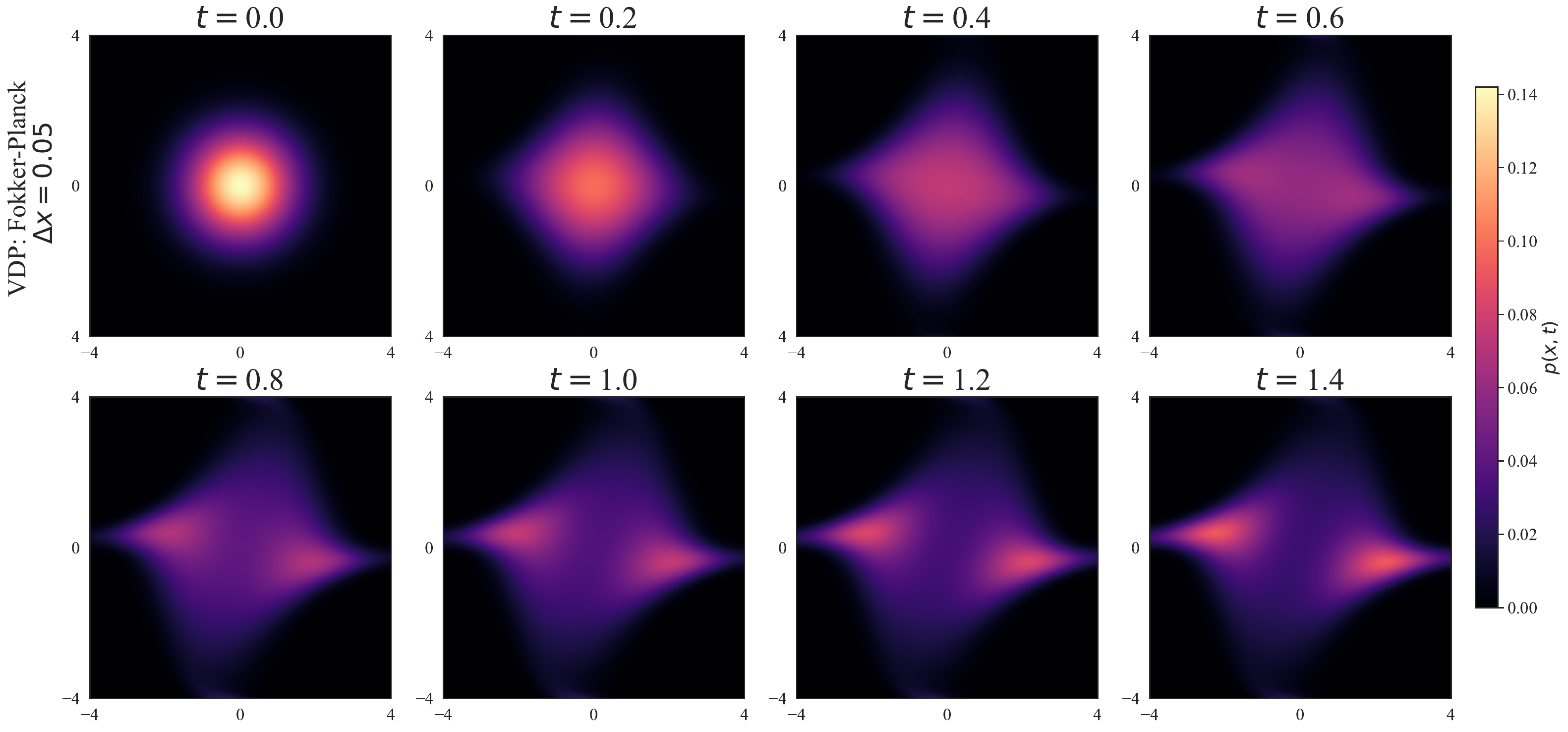}
    \caption{\textbf{Time-integration of the VDP FP equation.} Using the Crank-Nicholson scheme, we integrate the FP equation over time, using our discretised Laplacian. Starting from a Gaussian initial density centred at $(0,0)$ with standard deviation $\sigma =1$ in each direction, we step through with $\Delta t = 0.2$ up to $T=1.4$, we can see that the distribution equilibrates to the same distribution calculated directly in Fig. \ref{fig: VDPStationary}.}
    \label{fig: VDP FP}
\end{figure}

\begin{figure}
    \centering
    \includegraphics[width=\linewidth]{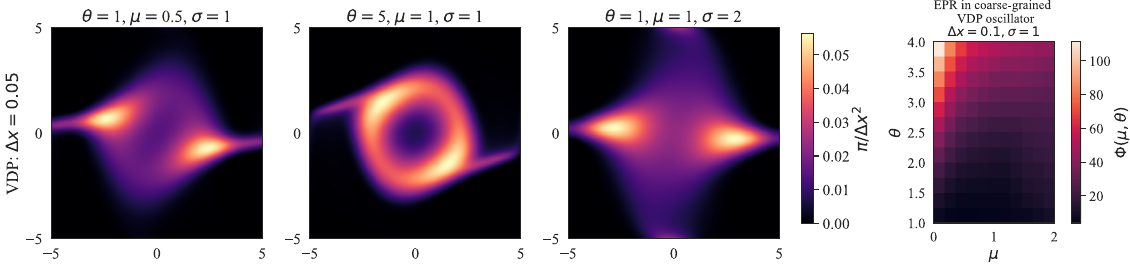}
    \caption{\textbf{NESS of the VDP oscillator}. The shape and nature of the stationary distribution shifts in response to changes in parameters. We illustrate this by calculating the stationary distribution of the VDP for an increased value of $\mu =3$, $\theta=5$ and $\sigma = 2$ respectively. An increased value of $\mu$ promotes nonlinearity in the oscillation, whilst $\theta$ promotes circular rotation around the origin. Finally, $\sigma$ increases the diffusivity of the process, leading to a less peaked distribution. Additionally, we can investigate how the EPR varies with $\mu$ and $\theta$ in the coarse-grained dynamics. We find that $\theta$ increases the EPR whilst $\mu$ decreases it.}
    \label{fig: VDP NESS}
\end{figure}

The \textit{van der Pol oscillator} (VDP) is a nonlinear, non-conservative oscillator that converges to an asymmetric limit-cycle in the absence of noise (see Fig. \ref{fig: VDPStationary}) \cite{strogatz2008nonlinear}. We consider a modified, stochastic VDP of the form
\begin{align}
    dx(t)& = \theta y \; dt + \sigma \;dW_x(t),\\
    dy(t)&= -\theta x + \mu y(1-x^2)\;dt + \sigma \;dW_y(t),\notag
\end{align}
where $\mu$ is a parameter that controls the nonlinearity and damping. The typical form of the VDP takes $\theta =1$, which has been added here as a free parameter to control the level of irreversible rotation in the drift field. There is no closed form for the stationary distribution of the VDP, thus we investigate it numerically with our coarse-grained dynamics.\\
\\
We begin by calculating the stationary distribution for the typical VDP with $\theta = 1$ and $\mu = 2$ for a range of decreasing step-sizes, shown in Fig. \ref{fig: VDPStationary}. We can see that, even for very coarse dynamics, the limit cycle behaviour is present. Moreover, unlike the Hopf oscillator, rotation around the stationary density does not occur uniformly quickly, thus the distribution is not rotationally invariant around the distribution, with a build up of probability along the apexes of the cycle.\\
\\
For models without an exact stationary density, it is worth additionally verifying that this is the stationary behaviour seen in the long-time-limit, by numerically integrating the FP equation directly. Using the \textit{Crank-Nicholson} scheme
\begin{align}
    P_{t+1} = B^{-1}AP_t, 
\end{align}
where $B= I - \frac{\Delta t}{2}L$ and $A=I + \frac{\Delta t}{2}L$ for a time-step $\Delta t$ \cite{LeVeque2007finitedifference}. We begin with an initial Gaussian density with zero mean and isotropic deviation $\sigma =1$ and integrate up to $T=1.4$ with $\Delta t = 0.2$ and $\Delta x = 0.05$. As shown in Fig. \ref{fig: VDP FP}, we see that the FP solution converges to the stationary distribution calculated directly.\\
\\
The shape and nature of the stationary distribution shifts with variations in parameters. In Fig. \ref{fig: VDP NESS}, we show three different parameter settings. Increasing $\mu$ increases the nonlinearity of the oscillation spreading the distribution over a larger limit cycle, whilst increased $\theta$ promotes circular rotation around the origin leading to a concentration of trajectories near the sinusoidal oscillation. Increasing $\sigma$ increases the diffusivity of trajectories and leads to a wider distribution.
Finally, we use our coarse-grained dynamics to investigate the behaviour of the EPR as a function of $\mu$ and $\theta$. We sweep between $\mu \in [0,2]$ and $\theta \in [1,4]$ as shown in Fig. \ref{fig: VDP NESS}. The EPR decreases as a function of $\mu$, which promotes nonlinearity in the oscillations. On the other hand, the EPR increases as a function of $\theta$, which promotes irreversible, circular rotation in the drift field.
\subsubsection{A pair of frustrated Kuramoto oscillators}
\begin{figure}
    \centering
    \includegraphics[width=\linewidth]{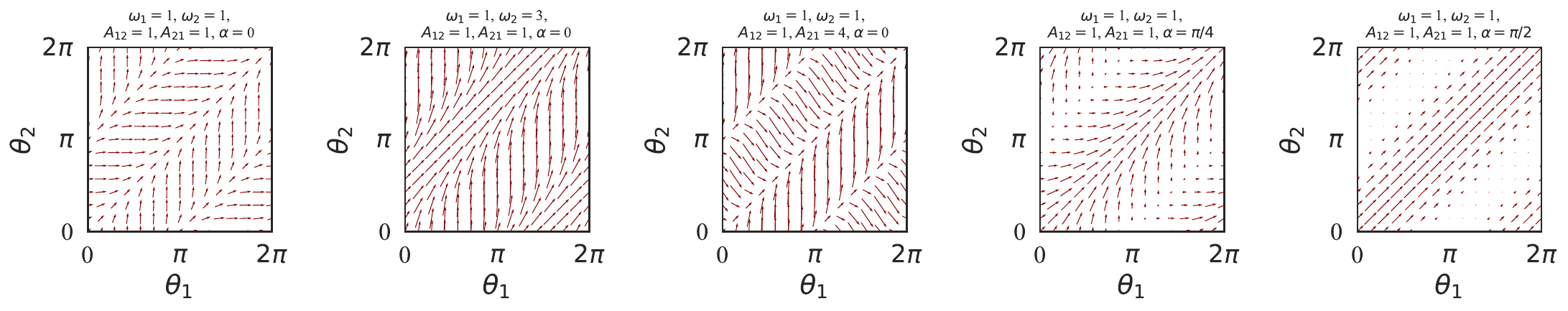}
    \caption{\textbf{Drift field of coupled Kuramoto oscillators.} We consider the drift field of a pair of Kuramoto oscillators. When the model is symmetric, the synchronised state is the only stable equilibrium, but this can be disrupted with asymmetry in the frequencies, interaction strengths or through frustration.}
    \label{fig: kuramoto vf}
\end{figure}

\begin{figure}
    \centering
    \includegraphics[width=\linewidth]{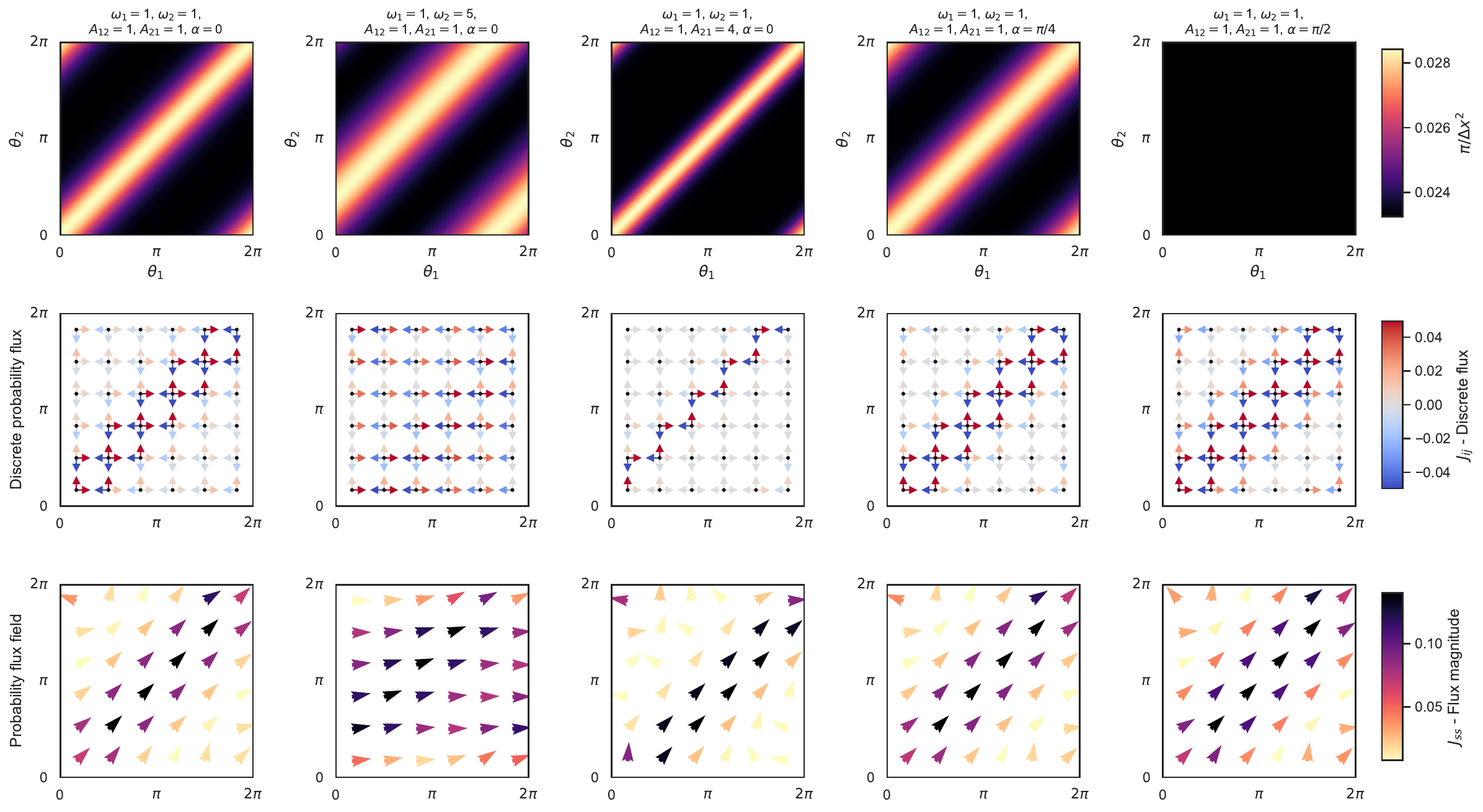}
    \caption{\textbf{NESS in the Kuramoto model}. For a range of different parameter values, we can compute the stationary density, discrete flux and continuous flux field in the NESS. We can see that the coupled Kuramoto system is in a NESS with most of the flux aligned along the synchronised oscillation. Nevertheless, the shape of the stationary distribution depends on the parameters.}
    \label{fig: kuramoto ness}
\end{figure}

The \textit{Kuramoto} oscillator is a model that describes the phase of interacting limit-cycle oscillators \cite{Kuramoto1984oscillations}, and is typically used to model synchronisation. It has found a range of applications describing neural dynamics \cite{nartallokalu2025review}, the flashing of fireflies, or the voltage oscillations of Josephson junctions \cite{strogatz2008nonlinear}.\\
\\
We first consider the case of a single, isolated Kuramoto oscillator
\begin{align}
    d\theta(t) = \omega \;dt + \sigma \;dW(t),
\end{align}
oscillating at a natural frequency of $\omega>0$ where $\theta(t)\in [0,2\bm{\pi})$. More explicitly, we arrive at the FP equation
\begin{align}
    \partial_tp&=-\omega\partial_\theta p + \frac{\sigma^2}{2}\partial^2_{\theta}p,
\end{align}
which yields the constant stationary density $\pi(\theta) = 1/2\bm{\pi}$, and stationary flux $\Jss(\theta) = \omega/2\bm{\pi}$ (Recall that we defined $\bm{\pi}$ to be the ratio of a circle's perimeter to its diameter, whilst $\pi$ is the stationary density). This leads to an EPR of $\Phi = 2\omega^2/\sigma^2$, from Eq. (\ref{eq: EPR FP}), but constraining the integral to the domain $[0,2\bm{\pi})$. This indicates that a single Kuramoto oscillator is in a NESS, with irreversibility driven by oscillation and mitigated by diffusion.\\
\\
Next we focus on a system with a pair of Kuramoto oscillators with additive noise. In addition, their interaction is \textit{frustrated}, a feature that can lead to nonequilibrium behaviour \cite{Gupta2014kuramoto}. We consider
\begin{align}
    d\theta_1(t)& = \omega_1 + A_{12}\sin(\theta_2-\theta_1+\alpha)\;dt + \sigma\;dW_1(t),\\
    d\theta_2(t)& = \omega_2 + A_{21}\sin(\theta_1-\theta_2+\alpha)\;dt + \sigma\;dW_2(t),\notag
\end{align}
where $\theta_i\in [0,2\bm{\pi})$ are the phases, $\omega_i>0$ are the natural frequencies, $\alpha$ is the frustration parameter, and $(A_{12}, A_{21})$ are the coupling strengths. Fig. \ref{fig: kuramoto vf} shows the drift field for a range of different values. In the symmetric case, we see that the synchronised state is a stable equilibrium, which can be disrupted with asymmetry in the natural frequencies, interaction strengths or through frustration.\footnote{In fact, for $A_{12}=A_{21}$ and $\alpha = \pi/2$, the NESS is explicitly solvable. The stationary distribution is uniform, $\pi = 1/4\bm{\pi}^2$ and the EPR is given by $\Phi = 2(\omega_1^2+\omega_2^2+A_{12}^2)/\sigma^2$. This shows that frustration can lead to a NESS, even in the case of symmetric coupling and no internal oscillations.}\\
\\
When coarse-graining this Kuramoto model with discrete states, we must enforce the periodic boundaries into our dynamics. Unlike the `no-flux' boundaries that we considered for the other models, we now enforce a periodic boundary on the discrete lattice such that it approximates $\mathbb{T}^2 = [0,2\bm{\pi})^2$. Our coarse-grained dynamics still preserve probability and therefore still yield a valid ME. Fig. \ref{fig: kuramoto ness} shows the stationary distribution, discrete probability flux, and continuous probability flux field for the Kuramoto model in a range of NESS. The parameters can shift the shape and nature of the stationary distribution, and both the density and flux are predominantly aligned with the synchronised oscillation.\\
\\
Finally, we investigate the EPR as a function of the various parameters.
\begin{figure}
    \centering
    \includegraphics[width=\linewidth]{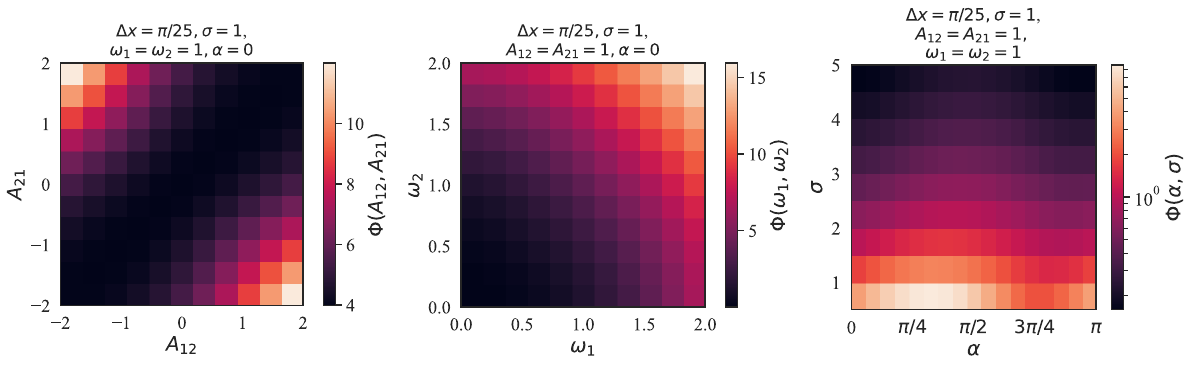}
    \caption{\textbf{EPR in coarse-grained Kuramoto oscillators.} We investigate how the EPR varies as a function of the parameters. We find that coupling asymmetry increases the EPR, whilst the absolute value of the natural frequencies drives the EPR, regardless of asymmetry. Finally, we find that the EPR-maximising value of the frustration depends on the noise-intensity.}
    \label{fig: kur num}
\end{figure}
In the first panel of Fig. \ref{fig: kur num}, we calculate the EPR as a function of various values of the coupling strengths. Whilst the system remains out of equilibrium for symmetric coupling, due to the internal oscillations, asymmetry in coupling strength drives the system further from equilibrium, in keeping with previous results on non-reciprocity in collective dynamics \cite{fruchart2021nonreciprocal,nartallokaluarachchi2024broken}. On the other hand, from the second panel, we see that only the absolute value, not the asymmetry, of natural frequencies drives the system from equilibrium, by increasing the speed of the collective oscillation. Finally, in the third panel, we investigate the effect of frustration and noise. Unsurprisingly, noise reduces the EPR by increasing reversible motion. Interestingly, the effect of frustration does not appear to be constant with respect to noise, with the EPR-maximising value of $\alpha$ varying slightly with the noise intensity, $\sigma$.

\section{Statistical inference for discrete-state processes}
\label{sec: inference}
When analysing NESS in physical and biological systems, we typically do not have access to the function forms of the drift or diffusion. Instead, we must analyse a diffusive process from observations of stochastic trajectories. In order to analyse the observed trajectories of a diffusion, it is common to infer a stochastic model. Whilst some methods attempt to directly infer a SDE \cite{Frishman2020stochasticforce,Friedrich2011complexity,bruckner2020inferring,elbeheiry2015inferencemap}, we will instead focus on the inference of a discrete-state Markov process from coarse-grained observations, which is relevant to a number of applied methods \cite{battle2016brokendetailedbalance,lynn2021detailedbalance,nartallokaluarachchi2024decomposing,Paijmans2017kai,Kimmel2018cellstate}. In this case, data takes the form of a sequence of coarse-grained states, but bifurcates into two possible situations.\\
\\
In the first, we have information about $K$ `jumps' and data takes the form, $\{(X_k,t_k):k\in\{1,...,K\}\}$, where $X_k$ is the state the process jumped into at `jump time' $t_k$. Given this knowledge about the exact jump times, the \textit{maximum likelihood estimator} (MLE) of the transition rate from state $j$ to state $i$ is given by
\begin{align}
    \hat{L}_{ij} & = \frac{N_{j\rightarrow i}}{T_j},\label{eq: MLE ctmc}
\end{align}
where $N_{j\rightarrow i}$ is the observed number of jumps from state $j$ to state $i$ that occurred during the $K$ steps, and $T_j$ is the \textit{holding time} i.e. the total time the process spent in state $j$ during the trajectory \cite{Bladt2005statisticalinference}.\\
\\
In the second, more typical, case, data comes in the form $\{X_t:t = k\Delta t, k\in\{1,...,K\}\}$, where we observe the process at a series of discrete snapshots, typically equispaced. Given such observations in discrete time, Eq. (\ref{eq: MLE ctmc}) is no longer the MLE of the transition rate of the CTMC, but instead an approximation whose accuracy depends on the time-step. Instead, we must infer a \textit{discrete-time Markov chain} (DTMC). In discrete time, the MLE of the \textit{transition probability matrix}\footnote{A TPM, $P$, defines a DTMC, where the entry $P_{ij}$ is the probability that the process will transition from state $j$ to state $i$ in a single time-step \cite{Ross2019probmodels}.} (TPM), $P$, is given by
\begin{align}
    \hat{P}_{ij} & = \frac{N_{j\rightarrow i}}{\sum_k N_{j\rightarrow k}},\label{eq: MLE dtmc}
\end{align}
where $N_{j\rightarrow k}$ is the number of observed transitions from state $j$ to state $k$\cite{Bladt2005statisticalinference}.\footnote{In practice, we must also avoid singularities, thus we employ a Bayesian prior. This means we add a `pseudo-count' for all valid transitions \cite{Sahasrabuddhe2025concise}.}\\
\\
The inference of a CTMC from $P$ is known as the \textit{embedding problem} for Markov matrices \cite{Casanellas2023embedding}. When monitoring a CTMC, with Laplacian $L$, at discrete time-steps, with interval $\Delta t$, the observation is a DTMC with TPM
\begin{align}
    P = \exp(L\Delta t). \label{eq: embedding}
\end{align}
Given $P$, the embedding problem is to find $L$ such that Eq. (\ref{eq: embedding}) is satisfied \cite{Casanellas2023embedding}.\footnote{In our case, $P=\hat{P}$ is calculated from the observations using Eq. (\ref{eq: MLE dtmc}).} Importantly, if $L$ solves the embedding problem for $\hat{P}$, then it is the MLE for the CTMC. However, it need not be unique as the matrix exponential is not injective. One simple condition that is sufficient for the existence and uniqueness of $L$ is
\begin{align}
    \inf_{i}(P_{ii}) \geq \frac{1}{2}.
\end{align}
A looser condition is that
\begin{align}
    \inf_{i}(P_{ii}) \cdot \det(P) > \exp(-\bm{\pi})\prod_iP_{ii},
\end{align}
where $L$ can be calculated using the unique logarithm of $P$ (see Ref. \cite{Bladt2005statisticalinference} and references therein). The embedding problem has significant implications for applications with coarse temporal observations including in sociology \cite{Singer1976representation}, economics \cite{Geweke1986mobility}, and evolutionary biology \cite{Verbyla2013embedding}, where direct \textit{expectation maximisation} algorithms are a sensible approach \cite{Bladt2005statisticalinference}. In the case of a diffusion and its coarse-graining, temporal granularity is typically high, thus we opt to use the approximate MLE of Eq. (\ref{eq: MLE ctmc}), which proved more numerically robust than computing the matrix logarithm.\\
\\
{\color{red} When inferring the transition rates with Eq. (\ref{eq: MLE ctmc}), we only count transitions between adjacent states, discarding transitions between non-adjacent states. In practice, we choose time and spatial resolution parameters such that at least 70\% of observed transitions are valid. In App. \ref{app: opt}, we consider an alternative optimisation-based approach to the embedding problem, which uses all transitions, but show that the results are comparable, whilst the method is significantly less efficient.}

\subsection{Numerical estimation of the EPR}
\begin{figure}
    \centering
    \includegraphics[width=\linewidth]{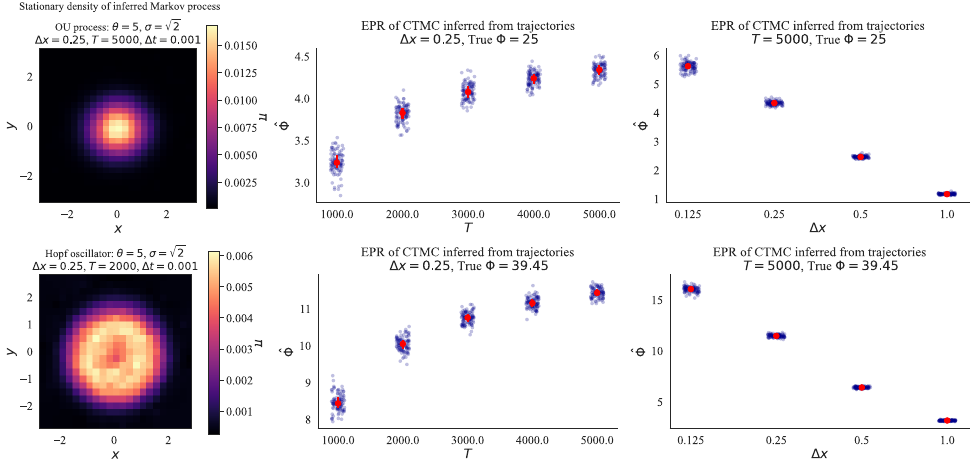}
    \caption{\textbf{Inferring NESS with Markov chain approximations.} Using samples from the OU process in Eq. (\ref{eq: OU ex}), we infer a CTMC and calculate the stationary distribution, which is a good approximation of the true density. We also compute the EPR for trajectories of different lengths and using different grid-sizes. We find that using longer trajectories and smaller grids leads to higher accuracy, but that the inferred EPR is substantially lower than the true value.}
    \label{fig: EPR_MLE_OU}
\end{figure}
We sample trajectories from the OU and HO processes (see App. \ref{app: sim OU} and \ref{app: sim Hopf} for sampling methods). Fig. \ref{fig: EPR_MLE_OU} shows the stationary distributions of the inferred CTMCs, which are a close approximation of the true density (see Sec. \ref{sec: approximation}). Moreover, we investigate the EPR estimates using trajectories of increasing length and decreasing grid-size. We find that longer trajectories and finer grid-sizes give more accurate estimates of the EPR, which are very stable over 100 iterations. However, in both cases, the inferred EPR is a substantial underestimate of the true EPR, highlighting the challenge of estimating the EPR from finite data, even for fine-grids.
\subsection{Determining if trajectories are from a NESS}
\begin{figure}
    \centering
    \includegraphics[width=0.5\linewidth]{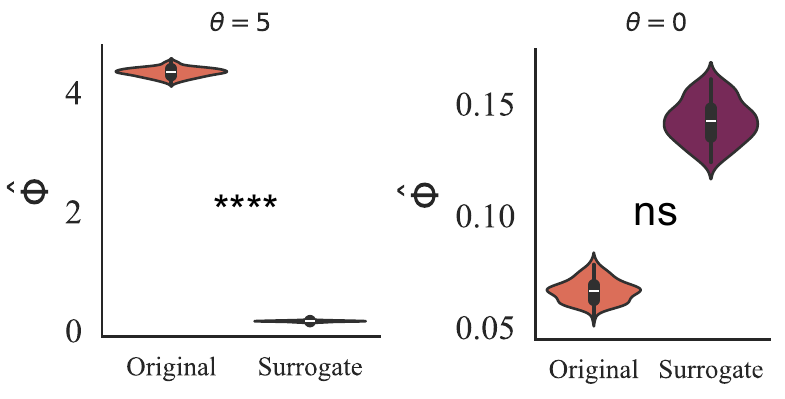}
    \caption{\textbf{Surrogate testing for NESS.} Using the shuffling procedure we can create surrogate Markov models from observed sequence data, and use this to perform testing to identify if real-world trajectories are from an ESS or a NESS. We illustrate this with trajectories from the OU process at $\theta = 5$, a NESS with $\Phi=25$, and $\theta = 0$, an ESS. Using a one-sided $t$-test, we find that at $\theta=5$ the process has significant (****, $p<0.0001$) entropy production, thus is in a NESS, whilst at $\theta = 0$, the process does not have significant entropy production, thus is not, as expected.}
    \label{fig: surrogates}
\end{figure}
A key drawback of inferring the EPR from sampled trajectories is that, by only considering finite data, we may obtain an unreliable estimate of the true EPR. In particular, we may naturally overestimate the irreversibility and flux in the stochastic process, and incorrectly determine that a reversible stationary process is out of equilibrium. For example, if we witness an odd number of transitions between two states, we will naturally observe an asymmetry, even if the true flux between the states is zero. As a result it is necessary to perform `surrogate testing' to determine if the measured irreversibility and EPR is a genuine statistical feature, or an artefact arising from the `noise-floor', the level of irreversibility that exists due to finite data. A typical approach to surrogate testing for time-irreversibility and entropy production is to shuffle trajectories in time, breaking the temporal order, and then re-inferring the EPR to obtain an estimate of the noise-floor \cite{battle2016brokendetailedbalance,lynn2021detailedbalance, nartallokalu2025review}. Whilst this approach works for some \textit{direct} measures of time-irreversibility, it is unsuitable for this Markov inference approach. This is because during the inference, we only consider transitions between adjacent cells - choosing the resolution of our grid such that this accounts for a sufficient number of time-points. When the trajectory is randomly shuffled, transitions between neighbouring cells become extremely rare, thus there are not enough valid transitions to infer the surrogate model.\\
\\
Instead, in order to test if the dynamics of the stochastic data are in fact from a NESS, we analyse the trajectory as before, with one crucial difference. For each valid transition in the trajectory, we flip the direction of the original transition with probability 1/2, thus erasing the net flux in the system that results in entropy production and irreversibility. Then, the EPR of the genuine trajectory can be compared to a surrogate distribution obtained from different realisations of the surrogate model. Under the null hypothesis that the genuine trajectory is from an ESS, we can test if the trajectory is from a NESS using a one-sided $t-$test. We illustrate this approach on trajectories from the OU process with $\theta=5$, an irreversible process with $\Phi = 25$, and $\theta = 0$, a reversible process. Fig. \ref{fig: surrogates} shows a comparison between the EPR of 100 trajectories sampled from the OU process compared with 100 surrogate models. At $\theta = 5$, we find that the EPR of the original trajectories is significantly higher (one-sided independent $t-$test; ****, $p<0.0001$), than the surrogate trajectories, thus we can conclude that the original trajectories are sampled from a NESS. On the other hand, at $\theta = 0$, we find that the original trajectories are not significantly higher than the surrogate models (one-sided independent $t-$test; ns, $p>0.05$), thus we must conclude, correctly, that the original process is reversible.\footnote{We note that in this case, the surrogate trajectories appear to have a significantly higher EPR than the original time-series. In theory, this shuffling approach should bring a NESS closer to detailed balance. On the other hand, if a trajectory is genuinely reversible, flipping a number of transitions is likely to \textit{increase} the EPR. Although this effect should dissipate as the number of transitions increases.}
\subsubsection{An application to real-world trajectories: group-polarisation in schooling fish}
\begin{figure}
    \centering
    \includegraphics[width=\linewidth]{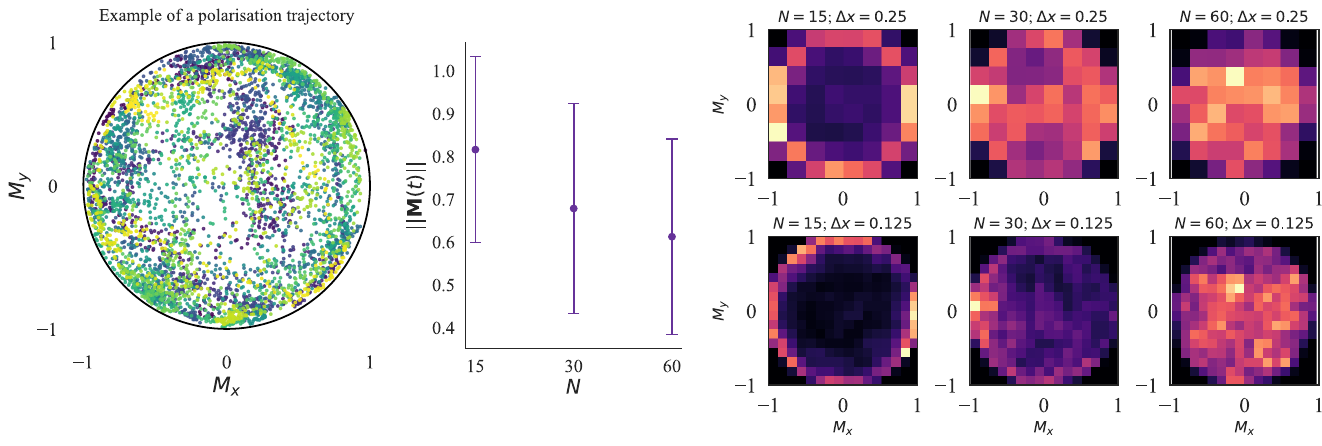}
    \caption{\textbf{Stationary dynamics in schooling fish.} We consider the group polarisation vector $\mathbf{M} = (M_x,M_y)$ for populations of schooling fish of size $N=15, 30$ and 60.  We find that the degree of collective alignment increases as group-size decreases, which can be seen by computing the mean and standard deviation of $||\mathbf{M}(t)||$ over time. Additionally, we can infer a discrete-state Markov process from the trajectories and calculate its stationary density, which again illustrates the higher level of collective alignment for smaller populations of fish.}
    \label{fig: fish1}
\end{figure}
\begin{figure}
    \centering
    \includegraphics[width=0.75\linewidth]{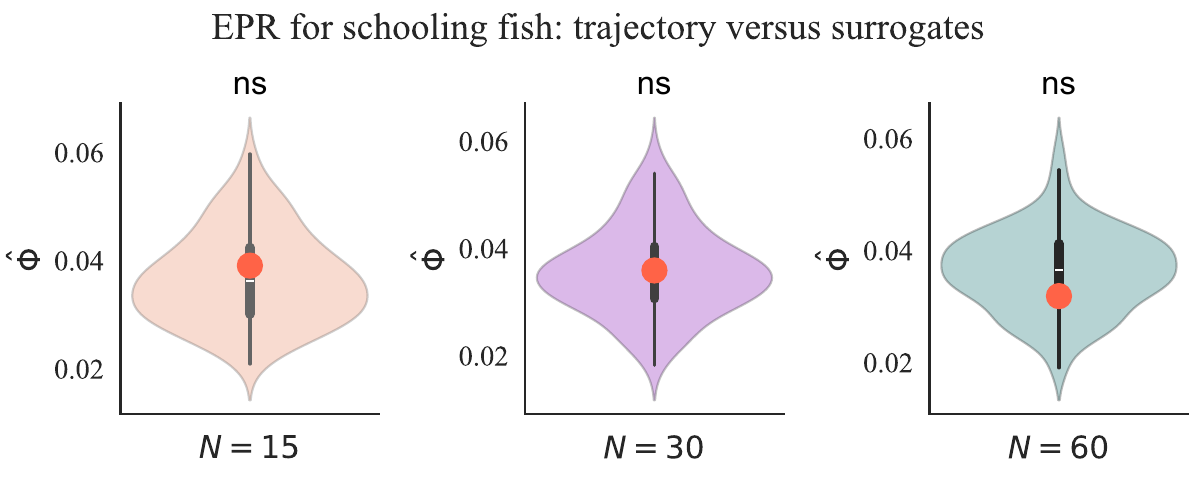}
    \caption{\textbf{EPR for populations of schooling fish compared to surrogate models.} Using the discrete-state model, we can measure the EPR of the trajectories of schooling fish. This is represented by the single orange dot in each panel. Using the surrogate model procedure, we obtain a `null' distribution (the violin plot) of the EPR in the surrogate model. We perform hypothesis testing to compare the single empirical trajectory with 100 surrogate models and find that the result is not significant across all population sizes.}
    \label{fig: fish epr}
\end{figure}

Finally, we apply this approach to a real-world case of a stationary diffusion, the polarisation trajectories of schooling fish, and determine that such a process is \textit{not} in a NESS. We follow the analysis of Jhawar et al. \cite{Jhawar2020schoolingfish}, who analyse populations of $N=15, 30$ and 60 fish moving in a tank. In particular, the positions, $\mathbf{x}_i(t)$, and velocities, $\mathbf{v}_i(t)$, of the fish can be extracted from video recordings. The \textit{group polarisation} is an order parameter describing their collective alignment given by
\begin{align}
    \mathbf{M}(t) = \frac{1}{N}\sum_{i=1}^N\hat{\mathbf{v}}_i(t),
\end{align}
where $\hat{\mathbf{v}}_i = \mathbf{v}_i/||\mathbf{v}_i||$, is the normalised velocity of fish $i$. A value of $||\mathbf{M}||$ close to 1 represents a coherent collective direction, whilst $||\mathbf{M}||\approx 0$ implies no collective direction, and isotropic individual moment. As $\mathbf{M}\in D_1(0) \subset \mathbb{R}^2$, by inferring an SDE, Jhawar et al. show that the group polarisation vector can be modelled as a stationary planar diffusion \cite{Jhawar2020schoolingfish}. Fig. \ref{fig: fish1} shows an example trajectory of the group polarisation vector. Moreover, using our discrete-state approach, we can infer the stationary densities for different grids and population sizes. In particular, the results shown in Fig. \ref{fig: fish1} align with the findings of Jhawar et al. indicating that `schooling', collective alignment, increases as population size decreases. This can be illustrated by both the mean value of $||\mathbf{M}(t)||$ over time, as well as the higher densities at the boundary of the unit disk (or its approximation with a square grid) in the steady-state solution of the inferred Markov chain.\\
\\
Next, we investigate if the stationary dynamics of the schooling fish are best described as a NESS or an ESS. For each of the population sizes, we infer the EPR of the original trajectory as well as the EPR of 100 surrogate models, obtained using the shuffling procedure. Fig. \ref{fig: fish epr} shows that, across population size, the EPR is not significantly higher for the original trajectories when compared to the surrogate model, from which we conclude that the process is not in a NESS. This result supports a well-known phenomenon in collective dynamics and active matter. Whilst collective dynamics are inherently \textit{active} at a local level \cite{toner1998flocks}, as they dissipate energy in order for agents to adjust their states dynamically, their collective behaviour can be described by equilibrium dynamics respecting time-reversal symmetry \cite{Fodor2016howfar} and maximum-entropy distributions \cite{Bialek2012stat,Mora2016local}.

\section{Discussion}
\label{sec: discussion}
Nonequilibrium steady-states are ubiquitous phenomena in physical and biological systems \cite{Gnesotto2018brokendetailedbalance,nartallokalu2025review,fang2019nonequilibrium}. Approximating the continuous-state dynamics of these diffusions with discrete-state models is a simple and effective way to analyse their dynamics, and has found applications to physical and biological data at multiple scales \cite{battle2016brokendetailedbalance,Li2019quantifying,lynn2021detailedbalance,nartallokaluarachchi2024decomposing}. Nevertheless, the consistency of the properties of the NESS between the continuous and discrete spaces was not well understood. Moreover, these approaches suffer the complex and obscurant effects of coarse-graining which can hide probability flux leading to underestimates of the entropy production rate \cite{Esposito2012coarsegraining} as well as induce non-Markovian effects that lead to artefactual scaling relationships \cite{Schwarz2024Memory}.\\
\\
In this paper, we presented a new approach to analytically derive a discrete-state Markov process from a stationary planar diffusion, using finite-volume methods. Moreover, we showed that the EPR of the approximation converges in the infinite-state limit, which we illustrated with both solvable and unsolvable examples of stationary diffusions. In the case of unsolvable processes, this approximation gives an accurate way to investigate the behaviour of the EPR under changes to the parameters, which we illustrated with the stochastic van der Pol and Kuramoto models. Finally, we considered the problem of statistical inference from trajectory data. Using sampled paths, our results show that discrete-state models significantly underestimate the EPR of a NESS, even for long trajectories and fine grids. Nevertheless, discrete-state models can be used to define a hypothesis test that can be used to determine if a trajectory is from a NESS or not, which we, again, illustrated with sampled trajectories. Finally, we considered an illustrative application to real-world trajectories by analysing the group-polarisation trajectories from schooling fish \cite{Jhawar2020schoolingfish}. 
\\\\
Our theoretical analysis and data-analytic methods  offer a much needed framework to investigate  the intersection and the consistency of discrete and continuous nonequilibrium stochastic processes. This approach has applications to both theoretical investigations of stochastic processes, as well as the data analysis of nonequilibrium stochastic trajectories.

\subsection*{Code and Data availability}
Python code, Jupyter notebooks, and trajectory data are available at \url{https://github.com/rnartallo/coarse_grained_diffusion}. The full dataset for the schooling fish has been made available by Jhawar et al. at \url{https://zenodo.org/records/3632470}.
\subsection*{Author contributions}
R.N.K. completed research and wrote the manuscript. R.L. and A.G. supervised research and edited the manuscript.
\subsection*{Acknowledgements}
The authors would like to thank Alexander Strang, Tassilo Schwarz and Jean-Charles Delvenne for important conversations about coarse-graining and stochastic processes. Additionally, they would like to thank Endre Süli, José Antonio Carrillo, Maria Bruna, {\color{red} and Jaime Agudo-Canalejo} for their advice on finite-volume discretisations. Finally, they would like to thank Jhawar et al. for making their data on schooling fish publicly available.\\\\
R.N.K acknowledge support in the form of an EPSRC Doctoral Scholarship from Grants No.
EP/T517811/1 and No. EP/R513295/1 and an Enrichment Community Award from The Alan Turing Institute. R.L.\ acknowledges support from the EPSRC grants EP/V013068/1, EP/V03474X/1 and EP/Y028872/1.
\bibliographystyle{IEEEtran} 
\bibliography{Bibliography}

@PREAMBLE{
 "\providecommand{\noopsort}[1]{}" 
 # "\providecommand{\singleletter}[1]{#1}%" 
}

@article{battle2016brokendetailedbalance,
    author = {Christopher Battle and Chase P. Broedersz and Nikta Fakhri and Veikko F. Geyer and Jonathon Howard and Christoph F. Schmidt and Fred C. Mackintosh},
    title = {Broken detailed balance at mesoscopic scales in active biological systems},
    journal = {Science},
    year = {2016},
    volume ={352},
    number ={6285},
    pages ={604--607}
}

@article{fruchart2021nonreciprocal,
    author = {Michel Fruchart and Ryo Hanai and Peter B. Littlewood and Vincenzo Vitelli},
    title = {Non-reciprocal phase transitions
},
    journal = {Nature},
    year = {2021},
    volume ={592},
    pages ={363--369}
}

@article{schnakenberg1976networktheory,
    author = {J.Schnakenberg},
    title = {Network theory of microscopic and macroscopic behavior of master equation systems
},
    journal = {Reviews of Modern Physics},
    year = {1976},
    volume ={48},
    number = {571}
}

@article{lax1960fluctuationsnonequilibrium,
    author = {Melvin Lax},
    title = {Fluctuations from the nonequilibrium steady state},
    journal = {Reviews of Modern Physics},
    year = {1960},
    volume ={32},
    number ={1}
}

@article{Masuda2017randomwalks,
  title={Random walks and diffusion on networks
},
  author={N. Masuda and M.A. Porter and R. Lambiotte
  },
  journal={Physics Reports},
  year={2017},
Volume ={716-717},
Pages ={1-58}
}

@article{lynn2021detailedbalance,
  title={Broken detailed balance and entropy production in the human brain},
  author={Lynn, Christopher W. and Cornblath, Eli J. and Papadopoulos, Lia and Bassett, Danielle S.},
  journal={Proceedings of the National Academy of Sciences},
  volume={118},
  number={47},
  year={2021},
  publisher={National Acad Sciences}
}

@article{martinez2019inferring,
    author = {Ignacio A. Martínez and Gili Bisker and Jordan M. Horowitz and Juan M. R. Parrondo
},
title = {Inferring broken detailed balance in the absence of observable currents
},
journal = {Nature Communications},
volume={10},
number ={3542},
year = {2019},
}

@article{seifert2019inference,
  title={From Stochastic
Thermodynamics to
Thermodynamic Inference
},
  author={Udo Seifert},
  journal={Annual Review of Condensed Matter Physics},
  year={2019},
  Volume={10},
    pages ={171-192}
}

@article{Schilling2022coarse,
  title={Coarse-grained modelling out of equilibrium},
  author={Tanja Schilling},
  journal={Physics Reports},
  volume={972},
  pages={1-45},
  year={2022}}

@article{Frishman2020stochasticforce,
    author = {Anna Frishman and Pierre Ronceray
},
title = {Learning Force Fields from Stochastic Trajectories
},
journal = {Physical Review X},
volume={10},
number ={021009},
year = {2020},
}

@article{DaCosta_2023,
year = {2023},
volume = {56},
number = {36},
pages = {365001},
author = {Lancelot Da Costa and Grigorios A Pavliotis},
title = {The entropy production of stationary diffusions},
journal = {Journal of Physics A: Mathematical and Theoretical}}

@article{fang2019nonequilibrium,
    author = {Xiaona Fang and Karsten Kruse and Ting Lu and Jin Wang},
    title = {Nonequilibrium physics in biology},
    journal = {Reviews of Modern Physics},
    year = {2019},
    volume ={91},
    number = {045004}
}

@article{Friedrich2011complexity,
    author = {Rudolf Friedrich and Joachim Peinke and Muhammad Sahimi and M. Reza Rahimi Tabar},
    title = {Approaching complexity by stochastic methods: From biological systems to turbulence},
    journal = {Physics Reports},
    year = {2011},
    volume ={506},
    number = {5},
    pages ={87--162}
}

@book{strogatz2008nonlinear,
    author = {Steven H Strogatz},
    title = {Nonlinear Dynamics And Chaos With Applications to Physics, Biology, Chemistry and Engineering},
    publisher = {Westview Press},
    year = {2008}
}

@phdthesis{strang2020applications,
  title        = {Applications of the {Helmholtz-H}odge decomposition to networks and random processes},
  author       = {Alexander Strang},
  year         = 2020,
  month        = {August},
  school       = {Case Western Reserve University},
  type         = {Ph{D} thesis}
}

@ARTICLE{Paige1982LSQR,
  author={Christopher C. Paige and Michael A. Saunders
 }, 
  title={{LSQR}: Sparse Linear Equations and Least Squares Problems
    }, 
journal ={ACM Transactions on Mathematical Software},
  year={1982},
  volume={8},
    number = {2},
    pages ={195-–209}}

@ARTICLE{vanKampen1981itostratonovich,
  author={N. G. van Kampen}, 
  title={Itô versus {S}tratonovich
    }, 
journal ={Journal of Statistical Physics},
  year={1981},
  volume={24},
    pages ={175–187}}

@ARTICLE{elbeheiry2015inferencemap,
  author={Mohamed El Beheiry and Maxime Dahan and Jean-Baptiste Masson
 }, 
  title={Inference{MAP}: mapping of single-molecule dynamics with {B}ayesian inference
}, 
journal ={Nature Methods},
  year={2015},
volume ={12},
pages ={594-595}}

@article{bruckner2020inferring,
    author = {David B. Br\"{u}ckner and Pierre Ronceray and Chase P. Broedersz
},
title = {Inferring the Dynamics of Underdamped Stochastic Systems
},
journal = {Physical Review Letters},
volume={125},
number ={058103},
year = {2020},
}

@ARTICLE{Yuan2017decomposition,
   author       = {Ruoshi Yuan and Ying Tang and Ping Ao},
   title = {{SDE} decomposition and {A-type} stochastic interpretation in nonequilibrium processes},
   year         = {2017},
   journal      = {Frontiers of Physics},
   volume       = {12},
   number = {120201}
}

@article{buckwar2022FHN,
    author = {Evelyn Buckwar and Adeline Samson and Massimiliano Tamborrino and Irene Tubikanec
},
    title = {A splitting method for {SDE}s with locally {Lipschitz} drift: Illustration on the {FitzHugh-Nagumo} model},
    journal = {Applied Numerical Mathematics},
    year = {2022},
    volume ={179},
    pages={191--220}
}

@article{brown1828brownian,
    author = {R. Brown},
    title = {A Brief Account of Microscopical Observations
Made in the Months of {June, July and August 1827}, on the
Particles Contained in the Pollen of Plants; and on the General Existence of Active Molecules in Organic and
Inorganic Bodies},
    journal = {Philosophical Magazine},
    year = {1828},
    volume ={4},
    number = {161}
}

@article{harrison2005stochasticneuronal,
    author = {L.M Harrison and O David and K.J Friston},
    title = {Stochastic models of neuronal dynamics},
    journal = {Philosophical Transactions of the Royal Society B},
    year = {2005},
    volume ={360},
    number = {1457},
    pages ={1075–1091}
}

@article{Li2011dicty,
    author = {Liang Li and Edward C Cox and Henrik Flyvbjerg
},
title = {{'Dicty dynamics'}: Dictyostelium motility as persistent random motion},
journal = {Physical Biology},
volume = {8},
number ={046006},
year = {2011},
}

@book{Franzke_O’Kane_2017, place={Cambridge}, author ={Christian L. E. Franzke and Terence J. O'Kane},title={Nonlinear and Stochastic Climate Dynamics}, publisher={Cambridge University Press}, year={2017}}

@article{Godreche2018OU,
  title={Characterising the nonequilibrium stationary states of {O}rnstein{-}{U}hlenbeck processes},
  author={Godr\`{e}che, C and Luck, JM
  },
  journal={Journal of Physics A: Mathematical and Theoretical},
  year={2018},
Volume ={52},
Number ={3}
}

@article{Bertini2015macroscopic,
    author = {Lorenzo Bertini and Alberto De Sole and Davide Gabrielli and Giovanni Jona-Lasinio and Claudio Landim},
    title = {Macroscopic fluctuation theory},
    journal = {Reviews of Modern Physics},
    year = {2015},
    volume = {87},
    number = {593}
}

@incollection{Touchette2013largedeviation,
    author = {Hugo Touchette and Rosemary J. Harris},
    title = {Large Deviation Approach to Nonequilibrium Systems},
    booktitle = {Nonequilibrium Statistical Physics of Small Systems: Fluctuation Relations and Beyond},
    publisher = {Wiley},
    year = {2013}
}

@article{Mielke2023saddle,
    author = {Alexander Mielke},
    title = {Non-equilibrium steady states as saddle points and {EDP} convergence for slow-fast gradient systems},
    journal = {Journal of Mathematical Physics},
    year = {2023},
    volume ={64},
    number = {123502}
}

@article{Kaiser2018canonical,
    author = {Marcus Kaiser and Robert L. Jack and Johannes Zimmer},
    title = {Canonical Structure and Orthogonality of Forces and Currents in Irreversible {M}arkov Chains},
    journal = {Journal of Statistical Physics},
    year = {2018},
    volume = {170},
    pages = {1019–1050}
}

@article{mielke2014gradientflows,
    author = {A. Mielke and M. A. Peletier and D. R. M. Renger},
    title = {On the Relation between Gradient Flows and the Large-Deviation Principle, with Applications to {M}arkov Chains and Diffusion},
    journal = {Potential Analysis},
    year = {2014},
    volume = {41},
    pages = {1293–1327}
}

@article{Mielke2016generalisation,
    author = {A. Mielke and D. R. M. Renger and M. A. Peletier},
    title = {A Generalization of {O}nsager’s Reciprocity Relations to Gradient Flows with Nonlinear Mobility},
    journal = {Journal of Non-equilibrium Thermodynamics},
    year = {2016},
    volume = {41},
    number = {2},
    pages = {141–149}
}

@article{Ayala2025reversibility,
    author = {Mario Ayala and Nicolas Dirr and Grigorios A. Pavliotis and Johannes Zimmer},
    title = {Reversibility, covariance and coarse-graining for {L}angevin dynamics: On the choice of multiplicative noise},
    journal = {arXiv},
    volume = {2511.03347},
    year = {2025}
}

@article{Heida2021finitevolume,
  title={Consistency and convergence for a family of finite volume discretizations of the {Fokker–Planck} operator},
  author={Martin Heida and Markus Kantner and Artur Stephan},
  journal={ESAIM: Mathematical Modelling and Numerical Analysis},
  year={2021},
  volume={55},
  pages ={3017–3042},
}

@article{Paijmans2017kai,
    author = {Joris Paijmans and David K. Lubensky and Pieter Rein ten Wolde},
    title = {A thermodynamically consistent model of the post-translational {Kai} circadian clock},
    journal = {PLOS Computational Biology},
    year = {2017},
    volume = {13},
    number = {3},
    pages = {e1005415}
}

@article{Kimmel2018cellstate,
    author = {Jacob C. Kimmel and Amy Y. Chang and Andrew S. Brack and Wallace F. Marshall},
    title = {Inferring cell state by quantitative motility analysis reveals a dynamic state system and broken detailed balance},
    journal = {PLOS Computational Biology},
    year = {2018},
    volume = {14},
    number = {1},
    pages = {e1005927}
}

@article{Holubec2019physically,
  title={Physically consistent numerical solver for time-dependent {Fokker-Planck} equations},
  author={Viktor Holubec and Klaus Kroy and Stefano Steffenoni},
  journal={Physical Review E},
  year={2019},
  volume={99},
  pages ={032117},
}

@article{nartallokaluarachchi2024broken,
  title={Broken detailed balance and entropy production in directed networks
},
  author={Ramón Nartallo-Kaluarachchi and Malbor Asllani and Gustavo Deco and Morten L. Kringelbach and Alain Goriely and Renaud Lambiotte},
   journal={Physical Review E},
  year={2024},
  volume={110},
    number = {034313}
}

@article{Wilson1983renormalisation,
    author = {Kenneth G. Wilson
},
    title = {The renormalization group and critical phenomena

},
    journal = {Reviews of Modern Physics},
    year = {1983},
    volume ={55},
    number ={583}
}

@article{nartallokalu2025review,
    author = {Ramon Nartallo-Kaluarachchi and Morten Kringelbach and Gustavo Deco and Renaud Lambiotte and Alain Goriely},
    title = {Nonequilibrium physics of brain dynamics},
    journal = {Physics Reports},
    volume = {1152},
    pages = {1-43},
    year = {2026}
}

@article{Schlichting2022Scharfetter,
    author = {André Schlichting and Christian Seis},
    title = {The {Scharfetter–Gummel} scheme for aggregation–diffusion equations},
    journal = {IMA Journal of Numerical Analysis},
    year = {2022},
    volume ={42},
    pages ={2361-2402}
}

@techreport{Frensley2004scharfettergummel,
    author = {William R. Frensley},
    title = {{Scharfetter-Gummel} Discretization Scheme
for Drift-Diffusion Equations},
    institution = {The University of Texas at Dallas},
    year = {2004}
}

@article{faccin2021stateaggregations,
    author = {Mauro Faccin and Michael T. Schaub and Jean-Charles Delvenne
},
    title = {State Aggregations in {M}arkov Chains and Block Models of Networks
},
    journal = {Physical Review Letters},
    year = {2021},
    volume = {127},
    number ={078301}
}

@ARTICLE{seifert2012thermodynamics,
  title={Stochastic thermodynamics, fluctuation theorems and molecular machines},
  author={Udo Seifert
  },
  journal={Reports on Progress in Physics},
  year={2012},
Volume={475},
Number ={12}
}

@BOOK{Schrodinger1944whatislife,
   author       = {E. Schrödinger},
   year         = 1944,
   title        = {What is Life? The Physical Aspect of the Living Cell and Mind},
   publisher    = {Cambridge University Press},
address = {Cambridge, United Kingdom}
}

@article{Esposito2012coarsegraining,
    author = {Massimiliano Esposito
},
    title = {Stochastic thermodynamics under coarse graining
},
    journal = {Physical Review E},
    year = {2012},
volume ={85},
number = {041125}
}

@article{busiello2019coarsegrained,
    author = {D M Busiello and J Hidalgo and A Maritan},
    title = {Entropy production for coarse-grained dynamics
},
    journal = {New Journal of Physics},
    volume ={21},
    number ={073004},
    year = {2019}
}

@article{rosvall2014memory,
    author = {Martin Rosvall and Alcides V. Esquivel and Andrea Lancichinetti and Jevin D. West and Renaud Lambiotte 
},
    title = {Memory in network flows and its effects on spreading dynamics and community detection
},
    journal = {Nature Communications},
volume={5},
number ={4630},
year = {2014},
}

@article{Marchetti2013hydrodynamics,
    author = {M. C. Marchetti and J. F. Joanny and S. Ramaswamy and T. B. Liverpool and J. Prost and Madan Rao and R. Aditi Simha},
    title = {Hydrodynamics of soft active matter},
    journal = {Reviews of Modern Physics},
    year = {2013},
    volume = {85},
    number = {1143}
}

@article{Ghosala2022inferring,
    author = {Aishani Ghosala  and  Gili Bisker},
    title = {Inferring entropy production rate from partially observed {L}angevin dynamics under coarse-graining},
    journal = {Physical Chemistry Chemical Physics},
    year = {2022},
    volume ={24},
    number = {39},
    pages ={24021-24031}
}

@article{Falasco2021localdetailed,
    author = {Gianmaria Falasco and Massimiliano Esposito},
    title = {Local detailed balance across scales: From diffusions to jump processes and beyond},
    journal = {Physical Review E},
    year = {2021},
    volume ={103},
    number = {042114}
}

@article{Gernert2014waiting,
    author = {Robert Gernert and Clive Emary and Sabine H. L. Klapp},
    title = {Waiting time distribution for continuous stochastic systems},
    journal = {Physical Review E},
    year = {2014},
    volume ={90},
    number = {062115}
}

@article{Meyberg2024entropy,
    author = {Ellen Meyberg and Julius Degünther and Udo Seifert},
    title = {Entropy production from waiting-time distributions for overdamped {L}angevin dynamics},
    journal = {Journal of Physics A: Mathematical and Theoretical},
    year = {2024},
    volume ={57},
    number = {25LT01}
}

@article{Skinner2021estimating,
    author = {Dominic J. Skinner and Jörn Dunkel},
    title = {Estimating Entropy Production from Waiting Time Distributions},
    journal = {Physical Review Letters},
    year = {2021},
    volume ={127},
    number = {198101}
}

@book{roldan2014thesis,
    author = {Edgar Roldán},
    title = {Irreversibility
and Dissipation
in Microscopic
Systems},
publisher ={Springer},
location = {New York},
    year = {2014}
}

@article{Gupta2014kuramoto,
    author = {Shamik Gupta and Alessandro Campa and Stefano Ruffo},
    title = {Kuramoto model of synchronization: equilibrium and nonequilibrium aspects},
    journal = {Journal of Statistical Mechanics: Theory and Experiment},
pages ={R08001},
    year = {2014}
}

@article{Gnesotto2018brokendetailedbalance,
    author = {F S Gnesotto and F Mura and J Gladrow and C P Broedersz},
    title = {Broken detailed balance and non-equilibrium dynamics in living systems: a review},
    journal = {Reports on Progress in Physics},
    year = {2018},
volume = {81},
number = {066601}
}

@article{Pareschi2018structure,
    author = {Lorenzo Pareschi and Mattia Zanella},
    title = {Structure Preserving Schemes for Nonlinear {Fokker–Planck} Equations and Applications},
    journal = {Journal of Scientific Computing},
volume ={74},
pages ={1575–1600},
    year = {2018}
}

@article{Li2019quantifying,
    title = {Quantifying dissipation using fluctuating currents
},
    author = {Junang Li and Jordan M. Horowitz and Todd R. Gingrich and Nikta Fakhri 
},
    journal = {Nature Communications},
volume ={10},
number ={1666},
    year = {2019}
}

@article{nartallokaluarachchi2024decomposing,
    title = {Decomposing force fields as flows on graphs reconstructed from stochastic trajectories
},
    author = {Ramón Nartallo-Kaluarachchi and Paul Expert and David Beers and Alexander Strang and Morten L. Kringelbach and Renaud Lambiotte and Alain Goriely
},
    journal = {Proceedings of the Third Learning on Graphs Conference (LoG 2024)},
volume ={PMLR 269
},
    year = {2024}
}

@article{Bladt2005statisticalinference,
  title={Statistical Inference for Discretely Observed {M}arkov Jump Processes
},
  author={Mogens Bladt and Michael Sørensen
},
  journal={Journal of the Royal Statistical Society Series B: Statistical Methodology},
  volume={67},
number ={3},
pages = {395–410},
  year={2005}
  }

@article{Casanellas2023embedding,
  title={The Embedding Problem For {M}arkov Matrices
},
  author={Marta Casanellas and Jesús Fernández-Sánchez and Jordi Roca-Lacostena
},
  journal={Publicacions Matemàtiques},
  volume={67},
number ={1},
  year={2023}
  }

@article{Verbyla2013embedding,
  title={The Embedding Problem for {M}arkov Models of Nucleotide Substitution
},
  author={Klara L. Verbyla and Von Bing Yap and Anuj Pahwa and Yunli Shao and Gavin A. Huttley 
},
  journal={PLOS One},
  volume={8},
number ={7},
pages = {e69187},
  year={2013}
  }

@article{Singer1976representation,
  title={The Representation of Social Processes by {M}arkov Models
},
  author={Burton Singer and Seymour Spilerman
},
  journal={American Journal of Sociology},
  volume={82},
number ={1},
pages = {1-54},
  year={1976}
  }

@article{Geweke1986mobility,
  title={Mobility Indices in Continuous Time {M}arkov Chains
},
  author={John Geweke and Robert C. Marshall and Gary A. Zarkin 
},
  journal={Econometrica},
  volume={54},
number ={6},
pages = {1407-1423},
  year={1986}
  }

@article{Matthews1990phasediagram,
  title={Phase Diagram for the Collective Behavior of Limit-Cycle Oscillators
},
  author={Paul C. Matthews and Steven H. Strogatz
},
  journal={Physical Review Letters},
  volume={65},
number ={14},
  year={1990}
  }

@article{Sheth2018hair,
  title={Nonequilibrium limit-cycle oscillators: Fluctuations in hair bundle dynamics
},
  author={Janaki Sheth and Sebastiaan W. F. Meenderink and Patricia M. Quiñones and Dolores Bozovic and Alex J. Levine},
  journal={Physical Review E},
  volume={97},
number ={062411},
  year={2018}
  }

@book{Kuramoto1984oscillations,
    author = {Yoshiki Kuramoto},
    title = {Chemical Oscillations, Waves, and Turbulence},
    publisher = {Springer-Verlag},
    year = {1984}
}

@book{Gardiner2010Methods,
    author = {Crispin Gardiner},
    title = {Stochastic Methods: A Handbook for the Natural and Social Sciences},
    publisher = {Springer},
    year = {2010}
}

@book{pavliotis2014stochproc,
    author = {Grigorios Pavliotis},
    title = {Stochastic Processes and Applications: Diffusion Processes, the Fokker-Planck and Langevin Equations},
    publisher = {Springer},
    year = {2014}
}

@book{Jiang2004noneq,
    title = {Mathematical Theory of Nonequilibrium Steady States: On the Frontier of Probability and Dynamical Systems},
    author = {Da-Quan Jiang and Min Qian and Min-Ping Qian
},
    publisher = {Springer},
    year = {2004}
}

@book{Ross2019probmodels,
    title = {Introduction to Probability Models},
    author = {Sheldon M. Ross},
    publisher = {Academic Press},
    year = {2019}
}

@article{toner1998flocks,
    author = {John Toner and Yuhai Tu},
    title = {Flocks, herds, and schools: A quantitative theory of flocking},
    journal = {Physical Review E},
    year = {1998},
    volume = {58},
    number = {4}
}

@article{Fodor2016howfar,
    author = {Étienne Fodor and Cesare Nardini and Michael E. Cates and Julien Tailleur and Paolo Visco and Frédéric van Wijland},
    title = {How Far from Equilibrium Is Active Matter?},
    journal = {Physical Review Letters},
    year = {2016},
    volume = {117},
    number = {038103}
}

@article{Mora2016local,
    author = {Thierry Mora and Aleksandra M. Walczak and Lorenzo Del Castello and Francesco Ginelli and Stefania Melillo and Leonardo Parisi and Massimiliano Viale and Andrea Cavagna and Irene Giardina 
},
    title = {Local equilibrium in bird flocks},
    journal = {Nature Physics},
    year = {2016},
    volume = {12},
    pages = {1153–1157}
}

@article{Bialek2012stat,
    author = {William Bialek and Andrea Cavagna and Irene Giardina and Thierry Mora and Edmondo Silvestri and Massimiliano Viale and Aleksandra M. Walczak
},
    title = {Statistical mechanics for natural flocks of birds},
    journal = {Proceedings of the National Academy of Sciences},
    year = {2012},
    volume = {109},
    number = {13},
    pages = {4786-4791}
}

@article{Sahasrabuddhe2025concise,
    author = {Rohit Sahasrabuddhe and Renaud Lambiotte and Martin Rosvall},
    title = {Concise network models of memory dynamics reveal explainable patterns in path data},
    journal = {Science Advances},
    year = {2025},
    volume = {11},
    number = {eadw4544}
}

@article{Schwarz2024Memory,
  title={Mind the memory: Consistent time reversal removes artefactual scaling of energy dissipation rate and provides more accurate and reliable thermodynamic inference
},
  author={Tassilo Schwarz and Anatoly B. Kolomeisky and Aljaž Godec},
  journal={arXiv},
  volume={2410.11819},
  year={2024}
  }

@book{LeVeque2007finitedifference,
    title = {Finite Difference Methods for Ordinary and Partial Differential Equations},
    author = {Randall J. LeVeque},
    publisher = {SIAM},
    year = {2007}
}

@article{Jhawar2020schoolingfish,
  title={Noise-induced schooling of fish
},
  author={Jitesh Jhawar and Richard G. Morris and U. R. Amith-Kumar and M. Danny Raj and Tim Rogers and Harikrishnan Rajendran and Vishwesha Guttal 
},
   journal={Nature Physics},
  year={2020},
  volume={16},
pages ={488-493}
}
\appendix
\section{The Scharfetter-Gummel discretisation}
\label{app: SG disc}

To derive the Scharfetter-Gummel discretisation in Eq. (\ref{eq: sg}), we follow the analysis of Ref. \cite{Frensley2004scharfettergummel}, extending to the 2D case with non-constant drift. We begin with the FVA of the FP equation
\begin{align}
    \frac{dp_{i,j}}{dt}& = -\frac{J^x_{i+\frac{1}{2},j}-J^x_{i-\frac{1}{2},j}}{\Delta x} - \frac{J^y_{i,j+\frac{1}{2}}-J^y_{i,j-\frac{1}{2}}}{\Delta y},
\end{align}
which is Eq. (\ref{eq: discrete FP}) of the main text. Next, we assume that the flux $J^{x}_{i+\frac{1}{2}}$, drift $f^{x}_{i+\frac{1}{2}}$ and diffusion $D^{x}_{i+\frac{1}{2}}$ are constant over the interval $[x_i,x_{i+1}]$. As a result, $p^j(x) :=p(x,y_j)$ satisfies the first-order ODE with constant coefficients
\begin{align}
    J^x_{i+\frac{1}{2},j}&= f^x_{i+\frac{1}{2},j}p^j(x) - D^x_{i+\frac{1}{2},j}\frac{dp^j}{dx}(x),
\end{align}
with boundary values $p^j(x_i)=p_{i,j}$ and $p^j(x_{i+1})=p_{i+1,j}$. For readability, we now drop the sub- and superscripts. We can solve this boundary value problem using an integrating factor,
\begin{align}
    Je^{-f(x-x_i)/D}& = \left(fp(x)-D\frac{dp}{dx}\right)e^{-f(x-x_i)/D},\\
    &=-D\frac{d}{dx}\left(e^{-f(x-x_i)/D}p(x)\right).
\end{align}
We then integrate this over the range $[x_i,x_{i+1}]$ where $\Delta x = x_{i+1}-x_i$. This yields
\begin{align}
    \int_{x_i}^{x_i+\Delta x} J e^{-f(x-x_i)/D}&= -D\left.\left(p(x)e^{-f(x-x_i)/D}\right)\right|_{x_i}^{x_{i+1}},\\
    J\frac{D}{f}\left(1-e^{-f\Delta x/D}\right)&=D\left(p_{i,j} - p_{i+1,j}e^{-f\Delta x/D}\right).
\end{align}
Reintroducing the subscripts, we obtain
\begin{align}
    J^x_{i+\frac{1}{2},j}& = \frac{f^x_{i+\frac{1}{2},j}\left(p_{i,j} - e^{-f^x_{i+\frac{1}{2},j}\Delta x/D^x_{i+\frac{1}{2},j}}p_{i+1,j}\right)}{1-e^{-f^x_{i+\frac{1}{2},j}\Delta x/D^x_{i+\frac{1}{2},j}}},
\end{align}
as in Eq. (\ref{eq: sg}). A similar derivation can be made for the flux terms $J^x_{i-\frac{1}{2},j}, J^y_{i,j+\frac{1}{2}}, J^y_{i,j-\frac{1}{2}}$. These can then be substituted into Eq. (\ref{eq: 2d FP}) where the coefficients of $p_{i+1,j}, p_{i-1,j}, p_{i,j+1}, p_{i,j-1}$ and $p_{i,j}$ can be derived.\\
\\
Whilst not reported here, we found that other structure-preserving schemes such as the \textit{Chang-Cooper} discretisation \cite{Pareschi2018structure} yielded valid master equations for these diffusions, whilst the centred-difference scheme only did so under step-size conditions. {\color{red} A number of structure-preserving schemes for the FP equation can be parametrised as a single family as described in Ref. \cite{Heida2021finitevolume}.}
\subsection{Boundary conditions}
\label{app : boundary}

The derivations of the transition rates in Sec. \ref{sec: markov approx} assume an infinite grid that approximates $\mathbb{R}^2$. In practice, this is not possible and we must instead enforce a finite boundary condition. There are three kinds of boundaries we can enforce with this grid-based set up: 1) absorbing 2) reflecting 3) periodic. In the first case, there is a net flux out of the grid, thus the process does not conserve probability. This creates significant problems when attempting to solve for the stationary distribution. In the second case, a reflective boundary implies `no-flux' at the border. This does preserve probability and is the most sensible choice for unbounded processes on $\mathbb{R}^2$. For a cell at the boundary, a transition (or two for the corner cells) is not possible, thus we must correct the diagonal entry of the Laplacian correspondingly in order to conserve probability. In all cases except the Kuramoto oscillator, we opt for this boundary condition. In the case of the Kuramoto oscillator, we have a periodic domain, thus we mirror this with a periodic boundary condition. This preserves probability and transitions are possible from one side of the grid to the other side. The transition rates must be amended accordingly. 

\subsection{Solving for the stationary distribution}
\label{app : solving stationary}
Solving for the stationary distribution involves solving the equation $L \pi = 0$. Whilst for a small number of discrete-states this can be calculated numerically in an efficient and exact way, for large state spaces (fine grids) this calculation becomes prohibitively slow. Instead, we solve
\begin{align}
    \min_{\pi: \sum_j \pi_j = 1}||L \pi||^2,
\end{align}
using LSQR \cite{Paige1982LSQR}, which capitalises on the sparsity of $L$. Whilst our theoretical derivations confirm that entries of $\pi$ should have the same sign, numerical instabilities may occur. For extremely fine grids, the discretisation error can be comparable to the error incurred through this approximation, thus convergence analysis becomes challenging.

{\color{red}
\section{Non-diagonal diffusion}
\label{app: nondiagonal}
In this paper we focus on the case of diagonal diffusion, and use the SG discretisation with a five-point stencil. This implies that the $dp_{i,j}/dt$ depends only on the four adjacent neighbours $\{p_{i+1,j}, p_{i-1,j}, p_{i,j+1}, p_{i,j-1}\}$. Here we will show that using a non-diagonal diffusion matrix requires a nine-point stencil, including the additional terms $\{p_{i+1,j+1}, p_{i-1,j+1}, p_{i+1,j-1}, p_{i-1,j-1}\}$. However, we show that this discretisation does not yield a valid master equation. Finally, we consider the use of shifted grids for coarse-graining processes which are diagonal under a coordinate transform, which includes the case of constant diffusion.

\subsection{Discretising non-diagonal diffusion with a nine-point stencil}

For simplicity, we will consider an SDE with no drift and non-diagonal, spatially-dependent diffusion, where $D(x,y)$ is given by
\begin{align}
    D(x,y) = \begin{pmatrix}
        D^x(x,y) & D^{xy}(x,y)\\
        D^{xy}(x,y) & D^y(x,y)
    \end{pmatrix},
\end{align}
where $D$ is positive definite. The FP flux is then given by $J = - \nabla \cdot (Dp) $. This process cannot be discretised with a generalisation of the SG discretisation due to these off-diagonal entries producing cross derivatives. We first note that
\begin{align}
    \nabla \cdot (Dp) & = \begin{pmatrix}
        \partial_x\left(D^xp\right) + \partial_y\left(D^{xy}p\right)\\
        \partial_x\left(D^{xy}p\right) + \partial_y\left(D^{y}p\right)
    \end{pmatrix}.
\end{align}
To obtain an expression for the discrete flux, we will use traditional centred differences \cite{LeVeque2007finitedifference}, for example we have
\begin{align}
    \partial_x\left(D^xp\right)|_{i+\frac{1}{2},j}&\approx \frac{D^x_{i+1,j}p_{i+1,j}-D^x_{i,j}p_{i,j}}{\Delta x},\\
    \partial_y\left(D^{xy}p\right)|_{i+\frac{1}{2},j}&\approx \frac{D^{xy}_{i+\frac{1}{2},j+1}p_{i+\frac{1}{2},j+1}-D^{xy}_{i+\frac{1}{2},j-1}p_{i+\frac{1}{2},j-1}}{2\Delta y},\\
    &\approx \frac{D^{xy}_{i+\frac{1}{2},j+1}\left(p_{i+1,j+1}+p_{i,j+1}\right)-D^{xy}_{i+\frac{1}{2},j-1}\left(p_{i+1,j-1}+p_{i,j-1}\right)}{4\Delta y}.
\end{align}
Repeating this calculation at each boundary, we can obtain the transition rates between states,
\begin{align}
    L_{(i,j)(i+1,j)}&= \frac{D^{xx}_{i+1,j}}{\Delta x^2} + \frac{D^{xy}_{i+1,j+\frac{1}{2}} - D^{xy}_{i+1,j-\frac{1}{2}}}{4 \Delta x \Delta y},\\
    L_{(i,j)(i-1,j)}&=  \frac{D^{xx}_{i-1,j}}{\Delta x^2} - \frac{D^{xy}_{i-1,j+\frac{1}{2}} - D^{xy}_{i-1,j-\frac{1}{2}}}{4 \Delta x \Delta y},\\
    L_{(i,j)(i,j+1)}&= \frac{D^{yy}_{i,j+1}}{\Delta y^2} + \frac{D^{xy}_{i+\frac{1}{2},j+1} - D^{xy}_{i-\frac{1}{2},j+1}}{4 \Delta x \Delta y},\\
    L_{(i,j)(i,j-1)}&= \frac{D^{yy}_{i,j-1}}{\Delta y^2} - \frac{D^{xy}_{i+\frac{1}{2},j-1} - D^{xy}_{i-\frac{1}{2},j-1}}{4 \Delta x \Delta y},\\
    L_{(i,j)(i+1,j+1)}&= \frac{D^{xy}_{i+\frac{1}{2},j+1} + D^{xy}_{i+1,j+\frac{1}{2}}}{4\Delta x \Delta y},\\
    L_{(i,j)(i-1,j+1)}&= -\frac{D^{xy}_{i-\frac{1}{2},j+1} + D^{xy}_{i-1,j+\frac{1}{2}}}{4\Delta x \Delta y},\\
    L_{(i,j)(i+1,j-1)}&= -\frac{D^{xy}_{i+\frac{1}{2},j-1} + D^{xy}_{i+1,j-\frac{1}{2}}}{4\Delta x \Delta y},\\
    L_{(i,j)(i-1,j-1)}&=\frac{D^{xy}_{i-\frac{1}{2},j-1} + D^{xy}_{i-1,j-\frac{1}{2}}}{4\Delta x \Delta y}.
\end{align}
Examining these rates we see that the positivity constraint is violated e.g. if $L_{(i,j)(i+1,j+1)}>0$ then $L_{(i+1,j)(i,j+1)}<0$. As a result, we cannot define a valid Markov chain with this rectangular grid unless $D^{xy}=0$.

\subsection{Diffusion which is diagonal under a coordinate transform}

Whilst we have showed that any process with non-diagonal diffusion does not yield a valid master equation, we now consider the special case of diffusion which is diagonal under a coordinate transform. Consider a process where the diffusion matrix can be written as,
\begin{align}
    D(x,y) & = V\Lambda(x,y)V^{-1},
\end{align}
where $V\in \mathbb{R}^{d\times d}$ is a matrix whose columns are the eigenvectors of $D$ and $\Lambda(x,y) = \text{diag}(\lambda_1(x,y),...,\lambda_d(x,y))$ is the diagonal matrix of eigenvalues i.e. the eigenvalues may change with space, but the eigenvectors remain constant. Clearly such a matrix-valued function can be diagonalised, $\Lambda(x,y) = V^{-1}D(x,y)V$, where the original approach can be applied in the transformed coordinates, which we call $(\nu,\eta)$ in $\mathbb{R}^2$. This is equivalent to constructing a grid of \textit{rotated rectangles} where the nodes are placed at
\begin{align}
    (x_i,y_j) = \mathbf{v}_1i\Delta\nu +\mathbf{v}_2j\Delta\eta,
\end{align}
where $\mathbf{v}_1, \mathbf{v}_2$ are the columns of $V$.

\paragraph{Example: Ornstein-Uhlenbeck process}
\begin{figure}
    \centering
    \includegraphics[width=0.8\linewidth]{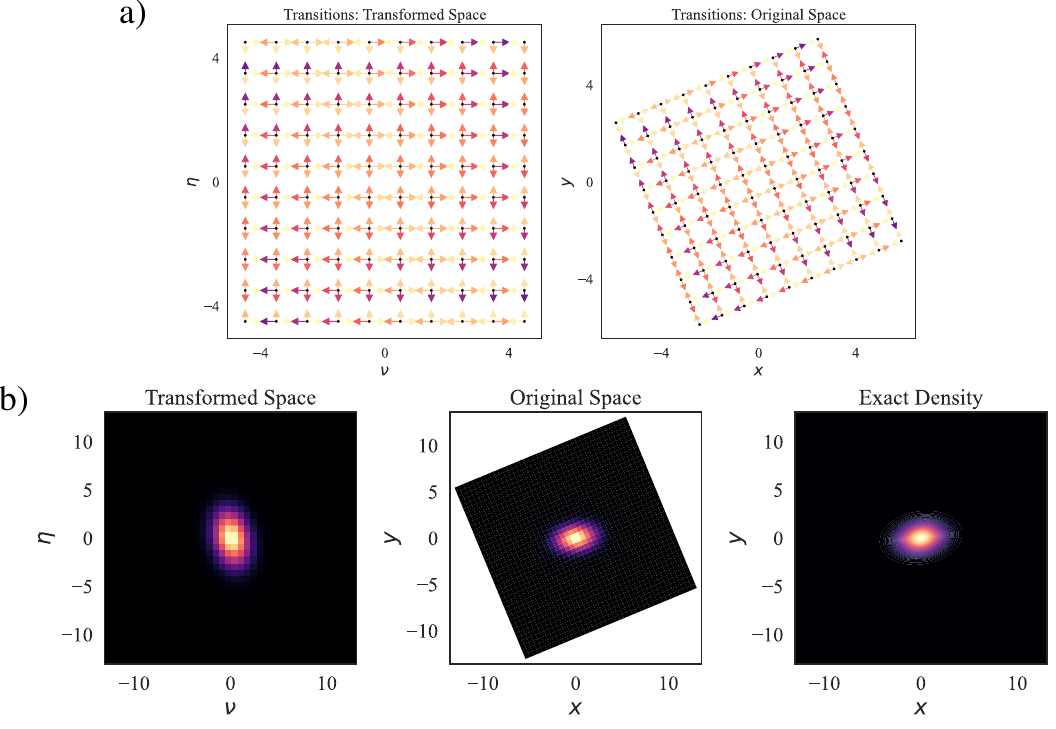}
    \caption{\textbf{Approximating processes with non-diagonal diffusion.} $a)$ We can apply the original approach, using a square grid, in the transformed space where the diffusion is diagonal. We can then return to the original coordinates where the grid is then rotated. $b)$ We illustrate this with the stationary distribution of the approximation and the exact density in both the transformed and original space.}
    \label{fig: rotated}
\end{figure}
We illustrate this coordinate transform approach on an OU process with
\begin{align}
    B &= \begin{pmatrix}
        2 & - 1\\
        1 & 2
    \end{pmatrix}, \hspace{10pt} D =\begin{pmatrix}
        4 & 1\\
        1 & 2
    \end{pmatrix},
\end{align}
i.e. where the diffusion is not diagonal, but diagonalisable and constant. We transform into the coordinate system where the diffusion is diagonal, and apply the SG discretisation with a square grid. The transition rates can then be considered as acting on a rotated grid in the original coordinates. Fig. \ref{fig: rotated} illustrates the transition rates (Panel $a)$), and stationary distribution (Panel $b)$) compared to the exact density in both the original and transformed coordinates.\\

For a diffusion matrix which does not have constant eigenvectors, it would be best approximated with an adaptive, irregular grid. However, this severely complicates both theoretical and numerical analysis.

\section{A variational approach to nonequilibrium steady-states}
\label{app: variational}

Here, we will briefly consider the \textit{energy dissipation principle}, which is central to variational approaches to NESS \cite{Bertini2015macroscopic,Touchette2013largedeviation,mielke2014gradientflows, Mielke2016generalisation, Mielke2023saddle, Kaiser2018canonical}.\\
\\
We begin with a diffusive process of the form in Eq. (\ref{eq: SDE}), and define the \textit{thermodynamic force},
\begin{align}
    F(x) & = \chi^{-1}(x) J(x) = D^{-1}(x)f_{\text{irr}}(x),
\end{align}
where $\chi(x) = D(x)p(x)$ is the \textit{mobility}. This force is the conjugate variable to the flux, allowing us to write the EPR at stationarity in the form $\Phi = \int J_{\text{ss}}\cdot F_{\text{ss}}\; dx$. Crucially, the variational structure is defined by two convex conjugate functionals. First, we define the \textit{dissipation potential} $\Psi(J)$, which captures the thermodynamic cost of observing a given fluctuation in the flux $J$ \cite{Bertini2015macroscopic},
\begin{align}
    \Psi(J) & = \frac{1}{2}\int J(x)^{\top} \chi^{-1}(x) J(x) \;dx.
\end{align}
Second, we define the \textit{dual dissipation potential}, $\Psi^*(F)$, which acts on the forces,
\begin{align}
    \Psi^*(F) & = \frac{1}{2}\int F(x)^{\top} \chi(x) F(x) \;dx = \frac{1}{2}\int p(x) F(x)^{\top} D(x) F(x) \;dx.
\end{align}
The central result of the energy dissipation principle is that the physical steady state minimizes the functional $\Psi(J) + \Psi^*(F) - \int J \cdot F dx$. Consequently, at the physical steady state (where $J_{\text{ss}} = \chi F_{\text{ss}}$), the total entropy production decomposes symmetrically into the sum of the potentials i.e. $\Phi = \Psi(J_{\text{ss}}) + \Psi^*(F_{\text{ss}})$. It is clear that the EPR has a \textit{quadratic} relationship with the flux, stemming from the quadratic nature of the dissipation potential.\\
\\
On the other hand, the dissipation potential of a CTMC is non-quadratic \cite{mielke2014gradientflows,Kaiser2018canonical}. To show this, we consider a CTMC and write the discrete flux, Eq. (\ref{eq: discrete flux}), as
\begin{align}
    J_{ij} & = \Lambda_{ij}\sinh\left(\frac{F_{ij}}{2}\right),
\end{align}
where $\Lambda_{ij} = 2\sqrt{L_{ij}\pi_jL_{ji}\pi_i}$ is the discrete analogue of the \textit{mobility}, and $F_{ij} = \log\left(L_{ji}\pi_i/L_{ij}\pi_j\right)$ is the discrete \textit{force} \cite{Kaiser2018canonical}. At stationarity, we have that $\Phi = \frac{1}{2}\sum_{i,j} J_{ij} F_{ij}$.\footnote{The factor of $\frac{1}{2}$ stems from double counting each directed edge $(i,j)$ which has the same contribution to the EPR.} To derive the dissipation potential and its conjugate, we note that $J_{ij} = \frac{\partial \Psi^*_{ij}}{\partial F_{ij}}$, thus
\begin{align}
    \Psi^*_{ij} & = 2\Lambda_{ij}\left(\cosh\left(\frac{F_{ij}}{2}\right)-1\right).
\end{align}
Via the Legendre transform, we have that
\begin{align}
    \Psi_{ij}(J_{ij}) = \sup_{F_{ij}} \left( J_{ij}F_{ij}- \Psi^*_{ij}(F_{ij})\right), 
\end{align}
which is achieved at $F_{ij}=2 \arcsinh\left(\frac{J_{ij}}{\Lambda_{ij}}\right)$, thus we have that\footnote{We use the identity $\cosh(\arcsinh(x)) = \sqrt{1+x^2}$.}
\begin{align}
    \Psi_{ij}(J_{ij}) & = 2J_{ij}\arcsinh\left(\frac{J_{ij}}{\Lambda_{ij}}\right) - 2\Lambda_{ij}\left(\sqrt{1+\frac{J_{ij}^2}{\Lambda_{ij}^2}} - 1\right).
\end{align}
The EPR along a directed edge $(i,j)$, $\Phi_{ij}$, can be written as the sum
\begin{align}
    \Phi_{ij} & = \Psi^*_{ij}(F_{ij}) + \Psi_{ij}(J_{ij}),\\
    & = 2 J_{ij}\arcsinh\left(\frac{J_{ij}}{\Lambda_{ij}}\right).
\end{align}
Clearly, this is function is not quadratic in the flux. However, in the continuum limit that we consider in Sec. \ref{sec: convergence}, the quantity $J_{ij}/\Lambda_{ij}\rightarrow 0$, as the rates increase. As a result, we can use the expansion $\arcsinh(x)\approx x +O(x^3)$, to obtain
\begin{align}
    \Phi_{ij} & \approx \frac{2 J_{ij}^2}{\Lambda_{ij}},
\end{align}
where the EPR is a quadratic function of the flux, normalised by the mobility, in direct analogy with the expression for a diffusive process.
}

\section{Sampling method for diffusions}
\subsection{Exact sampling for the Ornstein-Uhlenbeck process}
\label{app: sim OU}
In Sec. \ref{sec: inference}, we sample from the OU process. Unlike nonlinear processes, the OU process admits an exact simulation from its transition kernel \cite{DaCosta_2023}. Given an OU process
\begin{align}
dx(t) = -Bx(t) + \Sigma\;dW(t),
\end{align}
with $D = \frac{1}{2}\Sigma \Sigma^{\top}$, the solution is given by,
\begin{align}
    x(t) = e^{-{B}t}x(0) + \int_{0}^{t}e^{-B(t-s)}\Sigma \;dW(s),
\end{align}
Therefore, the transition kernel is given by
\begin{align}
   x(t) \sim \mathcal{N}( e^{-Bt}x(0),S(t)),
\end{align}
where,
\begin{align}
    S(t) &= \int_0^t2e^{-{B}(t-s)}{D}\left(e^{-{B}(t-s)}\right)^{\top}\;ds,
\end{align}
thus the process can be simulated exactly.
\subsection{Splitting method for the stochastic Hopf oscillator}
\label{app: sim Hopf}
In Sec. \ref{sec: inference}, we sample from the Hopf oscillator. This process does not admit exact simulation and we must employ numerical methods to simulate trajectories. We employ a geometric splitting integrator for the Hopf oscillator which preserves the underlying dynamics \cite{buckwar2022FHN}. Given an SDE,
\begin{align}
dx = f(x)\;dt + \Sigma \;dW(t),
\end{align}
we split the drift term into a nonlinear and linear part where we consider the additive noise alongside the linear drift,
\begin{align}
    dx^{[1]}(t)&=Ax^{[1]}(t) \;dt+ \Sigma\;dW(t),\\
    dx^{[2]}(t)&=G(x^{[2]}(t))\;dt,
\end{align}
where $f(x)=Ax + G(x)$. The linear part, ${x}^{[1]}$, defines an OU process which can be simulated exactly. If the splitting is chosen such that the ordinary differential equation for ${x}^{[2]}$ can be solved explicitly, then the solution of the full process can be approximated with the Strang approximation,
\begin{align}
    x(t+h) \approx \varphi_1^{[h/2]}\circ \varphi_2^{[h]} \circ \varphi_1^{[h/2]} (x(t)),
\end{align}
where $\varphi_i^{[h]}$ is a simulation from the exact $h$-time transition kernel of ${x}^{[i]}$.\\\\
For the Hopf oscillator, we use the decomposition,
\begin{align}
    d{x}^{[1]}(t)&=\begin{pmatrix}
 a&-\omega \\ 
\omega & a
\end{pmatrix}x^{[1]}(t)\;dt +\Sigma \;dW(t),\\
    dx^{[2]}(t)&=\begin{pmatrix}
-(x^2+y^2)x\\ 
-(x^2+y^2)y
\end{pmatrix}\;dt,
\end{align}
where $x^{[2]}= (x,y)$. The nonlinear part can be solved easily in polar coordinates.

{\color{red} 
\section{Statistical inference}
\subsection{Solving the embedding problem without discarding transitions}
\label{app: opt}

As mentioned, the embedding problem aims to find a Laplacian, $L$, whose observation in discrete-time has TPM,
\begin{align}
    P = \exp\left(L\Delta t \right).
\end{align}
Whilst $L$ is assumed to only admit certain transitions, $P$ can have non-zero entries in any position. We can formulate a solution to the embedding problem as a non-convex optimisation problem,
\begin{align}
\label{eq: optimisation}
    L^*= \argmin_{L\in \mathcal{L}}||P-\exp\left(L\Delta t\right)||_F,
\end{align}
where $\mathcal{L}$ is the set of valid Laplacian matrices which conserve probability, and only have non-zero transition rates for valid transitions. Due to presence of the matrix-exponential, this problem is non-convex.
\begin{figure}
    \centering
\includegraphics[width=0.85\linewidth]{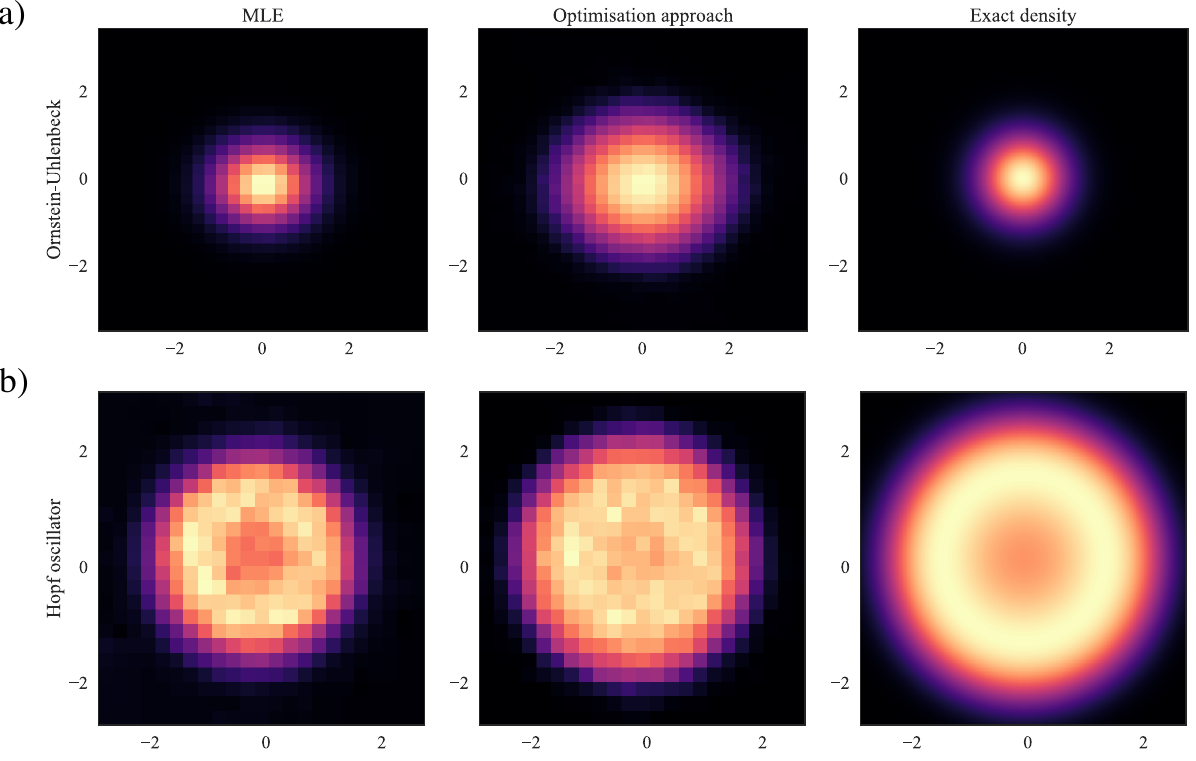}
    \caption{\textbf{Maximum-likelihood vs. optimisation for the embedding problem.} We compare the MLE estimator of Eq. (\ref{eq: MLE ctmc}) with the non-convex optimisation approach in Eq. (\ref{eq: optimisation}). We reconstruct the Laplacian matrix, and then solve for the stationary distribution, which we compare with the exact density for the $a)$ Ornstein-Uhlenbeck process, and $b)$ Hopf oscillator.}
    \label{fig: optim}
\end{figure}
We can compare this approach to the MLE in Eq. (\ref{eq: MLE ctmc}), infer the Laplacian, and solve for the stationary distribution. In practice, we use the standard \texttt{L-BFGS-B} method from \texttt{scipy.optimize}, but we find that this approach is far more inefficient than the direct estimation using Eq. (\ref{eq: MLE ctmc}).\\
\\
Fig. \ref{fig: optim} shows the result of a numerical experiment comparing the direct MLE with the optimisation-based approach for a trajectory from $a)$ the OU process, and $b)$ the Hopf oscillator, compared to the exact stationary density. We find that the methods yield comparably accurate stationary distributions.

\subsection{Taking the time-step to zero}
\label{app: timestep}

In Sec. \ref{sec: inference}, we show that as we increase the length of the stochastic trajectory, or increase the spatial resolution of the grid, our estimate of the EPR improves (increases), as shown in Fig. \ref{fig: EPR_MLE_OU}. It is tempting to assume that a similar behaviour will occur when we take the limit of the time-step in the stochastic trajectory to zero. However, as shown in Fig. \ref{fig: EPR_timestep}, this is not the case. Whilst we would expect convergence when taking both the time and spatial resolution to the limit, taking the time resolution to the limit whilst keeping the spatial resolution is a more complex scenario. It is true that as the time-step decreases initially, the transition rates are better estimated leading to an improved estimate of the EPR. However, beyond a certain point, this process begins to introduce increasingly prevalent non-Markovian effects. As a result our Markovian estimator becomes increasingly inaccurate.\\
\\
It is also worth noting that drift scales as $O(\Delta t)$, whilst diffusion scales as $O(\sqrt{\Delta t})$, thus, as the time-step decreases, reversible diffusion dominates irreversible drift leading to increased number of reversible transitions i.e. the per-step drift to diffusion ratio goes to zero. 
\begin{figure}
    \centering
    \includegraphics[width=0.5\linewidth]{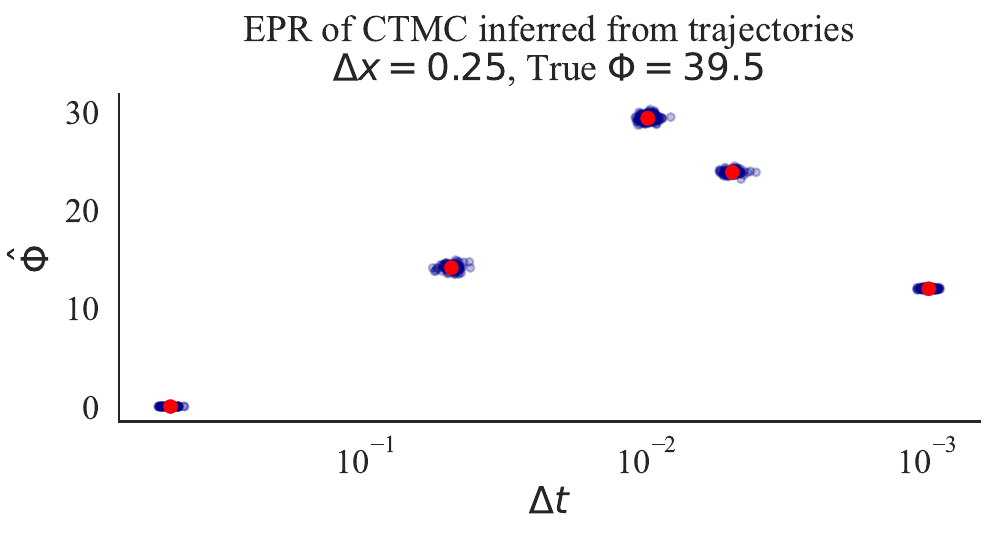}
    \caption{\textbf{EPR for decreasing time-steps.} We infer a CTMC from a trajectory sampled from the Hopf oscillator and measure the EPR, for a series of decreasing time-steps and a fixed spatial resolution. We find that the EPR does not converge as the drift to diffusion ratio goes to zero.}
    \label{fig: EPR_timestep}
\end{figure}
}
\end{document}